\documentclass[superscriptaddress,notitlepage,nofootinbib]{revtex4-2}

\usepackage{natbib}
\usepackage{physics}
\usepackage{xcolor}
\usepackage{amsmath, amsfonts}
\usepackage{siunitx}
\usepackage{graphicx}
\usepackage[english]{babel}
\usepackage[utf8]{inputenc}
\usepackage[autostyle]{csquotes}\MakeOuterQuote{"}
\usepackage[shortlabels]{enumitem}
\usepackage{dsfont}
\usepackage{comment}
\usepackage{verbatim}
\usepackage{xpatch}
\usepackage{tcolorbox}
\usepackage{float}
\usepackage{etoolbox}
\usepackage{amssymb}
\usepackage[hidelinks]{hyperref}	
\usepackage[capitalize]{cleveref}
\usepackage{soul}

\date{\today}

\definecolor{cadetgrey}{rgb}{0.57, 0.64, 0.69}
\newtheorem{result}{Result}

\newcommand{\set}[1]{\{#1\}}

\newcommand{\identity}{\mathds{1}}

\newcommand{\abit}{a}
\newcommand{\abasis}{\beta}
\newcommand{\aint}{I}

\newcommand{\SShi}{S}   %Shield
\newcommand{\STra}{T}   %Transmitted
\newcommand{\SAnc}{A}   %Ancilla

\begin{document}
\setstcolor{red}

\title{Security of practical modulator-free quantum key distribution}

\author{Álvaro Navarrete}
    \email{anavarrete@vqcc.uvigo.es}
	\affiliation{\affvqcc} \affiliation{\affuvigo} \affiliation{\affatlantic} 
\author{Víctor Zapatero}
    \affiliation{\affvqcc} \affiliation{\affuvigo} \affiliation{\affatlantic} 
\author{Marcos Curty}
    \affiliation{\affvqcc} \affiliation{\affuvigo} \affiliation{\affatlantic} 

\newcommand{\affvqcc}{Vigo Quantum Communication Center, University of Vigo, Vigo E-36310, Spain}
\newcommand{\affuvigo}{Escuela de Ingeniería de Telecomunicación, Department of Signal Theory and Communications, University of Vigo, Vigo E-36310, Spain}
\newcommand{\affatlantic}{atlanTTic Research Center, University of Vigo, Vigo E-36310, Spain}

\begin{abstract}
Recent advancements in quantum key distribution have led to the development of various modulator-free transmitters. Among their advantages, they offer enhanced security against Trojan-horse attacks.  However, practical implementations emit residual pulses that, while not used in the quantum communication, still carry information about Alice's state preparation process. While the intensity of these pulses is typically attenuated, the extinction ratio of practical intensity modulators is always finite, and therefore it remains crucial to account for the residual information leakage at the security-proof level. In this work, we prove the security of these transmitters in such setting and evaluate their performance. We find that the secret-key rate of the protocol is severely affected when the information leakage is not sufficiently small, which highlights the importance of taking into account this imperfection.
\end{abstract}

\maketitle

\section{Introduction}
The security of quantum key distribution (QKD)~\cite{lo2014secure,xu2020secure,pirandola2020advances} typically relies on the assumption that the physical devices employed in its operational implementation adhere to the theoretical models used in the security proofs. However, this assumption is very hard ---if not impossible--- to guarantee in practice, as imperfections are inevitable in real-world devices~\cite{zapatero2024implementation,marquardt2024implementation}. Moreover, an eavesdropper ---often referred to as Eve--- could interact with the components of a QKD system to manipulate their behavior or learn information about their internal settings, \textit{e.g.}, by launching a so-called Trojan-horse attack (THA) \cite{gisin2006trojan,vakhitov2001large,jain2014trojan,sajeed2017invisible,lucamarini2015practical,tamaki2016decoy,wang2018finite,navarrete2022improved}. In this latter type of attack, Eve injects light into Alice's transmitter and analyzes the light that is reflected, which is encoded by the optical modulators on its way back to the channel. In doing so, Eve creates a side channel that leaks information about the settings selected by Alice to encode each of the emitted signals.

In this context, different modulator-free transmitters have been recently proposed that root out this crucial vulnerability. These transmitters allow the preparation of quantum signals without the need for active modulators, thereby rendering them immune to THAs. A prominent example are the passive decoy-state QKD transmitters recently introduced in~\cite{zapatero2023fully,wang2023fully,zapatero2024finite,lu2023experimental,hu2023proof} based on the seminal works in~\cite{curty2009non,curty2010passive,curty2010passive2}. In these setups, phase-randomized coherent pulses are passively combined in a linear optics network to generate optical states with random intensity and phase/polarization. These states are subsequently post-selected by using a measurement apparatus within the transmitter to determine the final output quantum signals. Remarkably, these transmitters avoid the use of random number generators as well, in so notably simplifying the driving electronics. Another example is the modulator-free decoy-state QKD transmitter proposed in~\cite{lo2023simplified}, in which the phase and intensity of the output pulses is set by combining  optical injection locking techniques with direct electronic modulation of both the master and slave lasers, as well as coherent interference. This requires the use of random numbers to modulate the electronic modulators used to tune the laser directly, yet it provides immunity to THAs, as well as a compact configuration.

Importantly, however, practical implementations of this type of transmitters~\cite{lu2023experimental,hu2023proof,lo2023simplified} generate, together with the desired quantum signals, some additional optical pulses that contain sensitive information about their intensity and bit/basis settings. To prevent this, an intensity modulator (IM) is often placed at the output of the transmitter to block these unwanted pulses. The behavior of this IM is predetermined, \textit{i.e.}, it does not depend on Alice's settings, and so Eve cannot obtain any useful information tampering with it. Unfortunately, real-world IMs have a finite extinction ratio, and thus cannot perfectly remove the undesired pulses. Therefore, it is still crucial to account for the residual amount of information leakage in the security proof.

In this work, we prove the security of modulator-free decoy-state QKD transmitters and evaluate their performance against this kind of inherent information leakage present in practical settings. For this, we build on security-proof techniques that have been proven adequate to address the security of QKD against general attacks in scenarios where the transmitted states are imperfect~\cite{gottesman,Lo-Preskill,Lo-Ma,curras2023security}. We find that the performance of these protocols is severely affected when the information leakage is not sufficiently attenuated, which stresses the need to carefully account for this sort of vulnerability at the security-proof level.

The paper is structured as follows. In~\cref{sec:PassiveTransmitter} we analyze the security of the type of passive transmitters based on post-selection proposed in~\cite{lu2023experimental,hu2023proof,lo2023simplified} in the presence of information leakage. In~\cref{sec:OILtransmitter}, we adapt such security analysis for the modulator-free transmitter based on optical injection locking introduced in~\cite{lo2023simplified}. We evaluate the performance of both transmitters as a function of the information leaked in~\cref{sec:Results}. Finally, some conclusions are gathered in~\cref{sec:conclusions}. Additional information is provided in the Appendices.

\section{Passive Transmitter based on post-selection}\label{sec:PassiveTransmitter}
\begin{figure}
    \centering
    \includegraphics[width=0.8\columnwidth]{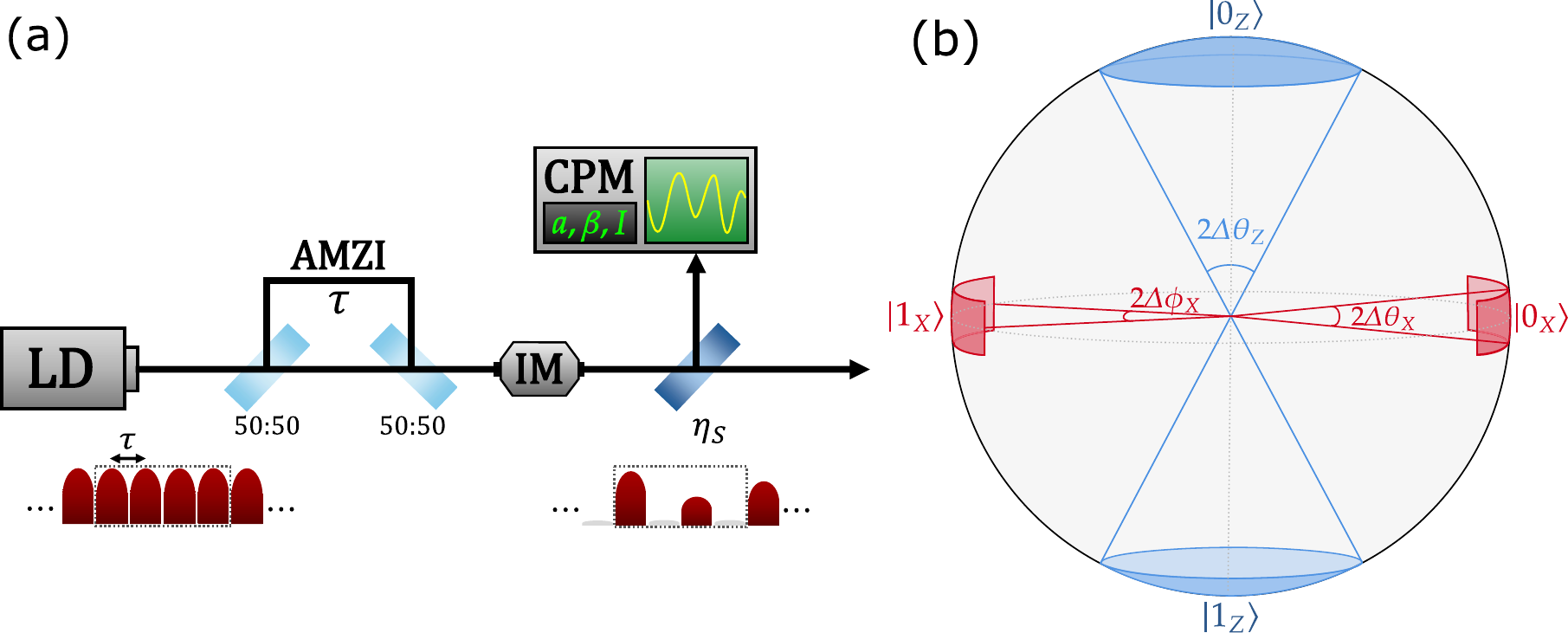}
    \caption{(a) Schematic representation of the passive transmitter proposed in~\cite{lu2023experimental}. The laser diode (LD) emits a train of phase-randomized coherent pulses that is fed into an asymmetric Mach-Zehnder interferometer (AMZI) with a time delay $\tau$ that matches the time separation between two consecutive pulses. At the output of the AMZI, an intensity modulator (IM) blocks the pulses corresponding to the odd time slots, and a beamsplitter with low transmittance $\eta_{S}$ distributes the signals between the classical post-selection module (CPM) ---which determines the values of the bit $a$, the basis $\beta$, and the intensity $I$ associated with each particular 4-pulse round--- and the quantum channel. (b) Post-selection regions for the target variables $\theta$ and $\phi$, which depend on the pre-fixed angles $\Delta\theta_Z$, $\Delta\phi_Z$ and $\Delta\phi_X$. \label{fig:scheme}}
\end{figure}

For illustration purposes, we focus first on the passive time-bin decoy-state BB84 transmitter presented in~\cite{lu2023experimental}, which is based on post-selection (see~\cref{fig:scheme}a). Importantly, we remark that the security analysis we introduce here is rather general, and can be applied to other passive transmitters based on post-selection as, \textit{e.g.}, those reported in~\cite{zapatero2023fully,wang2023fully,zapatero2024finite,lu2023experimental,hu2023proof}. Besides, for concreteness, we shall consider an asymmetric decoy-state BB84 protocol in which Alice and Bob extract key from a single basis, while they use the complementary basis for parameter estimation. The protocol is structured as follows:
\begin{enumerate}
    \item \textit{State preparation:} For each protocol round $u=1,\dots,N$, Alice prepares a quantum signal in a mixed state that is close to an ideal decoy-state BB84 signal, and sends it to Bob. Specifically, in each round,
    \begin{enumerate}
        \item Alice's laser diode emits a train of four consecutive coherent pulses $\bigotimes_{t=1}^{4}\ket{\sqrt{\mu_{\rm in}}e^{i\phi_t}}_t$, where the strong (classical level) intensity $\mu_{\rm in}$ is fixed and each phase $\phi_t$ is uniformly random.
        \item Such pulse train is fed into an asymmetric Mach-Zehnder interferometer (AMZI) with a delay $\tau$ that matches the time separation between consecutive pulses such that they interfere. After the AMZI, the state of the pulse train is $\bigotimes_{t=1}^{4}\ket*{\sqrt{\mu_{\rm in}}/2(e^{i\phi_t}+e^{i\phi_{t-1}})}_t$, where $\phi_{0}$ denotes the optical phase of the last pulse of the previous round (\textit{i.e.}, $\phi_{0}$ in the present round corresponds to the phase $\phi_{4}$ of the previous round).
        \item Next, an IM blocks the odd pulses (time bins 1 and 3). These pulses are not used to convey quantum information, and so they are blocked to prevent information leakage.
        \item Finally, the even pulses ---\textit{i.e.}, time bins 2 and 4, which we conveniently redefine as the early ($e$) and late ($l$) time bins--- are directed to an asymmetric beamsplitter with low transmittance $\eta_S$, which reflects most of the light to a classical post-selection module (CPM), while transmitting the remaining weak light to the quantum channel. The CPM either outputs the bit value $a\in\set{0,1}$, the basis $\beta\in\set{Z,X}$, and the intensity setting $I\in\set{I_0,I_1,I_2}$ associated with the round, or it outputs an inconclusive outcome.
    \end{enumerate} 
    \item \textit{Measurement:} Bob measures the incoming time-bin signals in the $Z$ or the $X$ basis with a predefined probability $p_{\beta_B}$, with $\beta_B\in\set{Z,X}$.
    \item \textit{Sifting:} Bob publicly announces the rounds in which he observes a detection, as well as their corresponding measurement basis. Alice records the data corresponding to the detected basis-match rounds and announces the post-selected bases $\beta$ and intensity settings $I$ associated to these rounds to Bob. The bit values of detected rounds where both parties selected the $Z$ basis and Alice selected the intensity setting $I_0$ are used to construct the sifted keys.
    \item \textit{Parameter estimation:} Alice and Bob reveal the bit values of the detected rounds where both parties selected the $X$ basis. If the statistics of these rounds satisfy certain constraints, they proceed with the following step. Otherwise, they abort the protocol.
    \item \textit{Post-processing:} Alice and Bob run standard error correction, error verification, and privacy amplification sub-routines to distill two identical secret keys with high probability.
    
\end{enumerate}

\subsection{Classical post-selection}
Here we describe the classical post-selection performed by the CPM. For convenience, we define the following random variables, which are functions of the random phases $\phi_1,\dots\phi_4$ that characterize the pulse train of any given round~\cite{lu2023experimental}:
\begin{align}\label{eq:transformations1}
    \mu &:=\mu_e+\mu_l, & \mu_e&:=\frac{\mu_{\rm max}}{2}[1+\cos(\phi_1-\phi_2)], & \mu_l&:=\frac{\mu_{\rm max}}{2}[1+\cos(\phi_3-\phi_4)],\nonumber\\
    \phi &:=\phi_l-\phi_e, & \phi_e&:=\frac{1}{2}(\phi_1+\phi_2)+ \varphi_{\phi_1-\phi_2}, & \phi_l&:=\frac{1}{2}(\phi_3+\phi_4)+ \varphi_{\phi_3-\phi_4},\\
    \theta &:=2 \arccos(\sqrt{\frac{\mu_e}{\mu}})\nonumber,
\end{align}
where $\mu$ represents the intensity in the joint mode $el$ comprising the early and late optical modes $e$ and $l$, $\mu_e$ ($\mu_l$) denotes the intensity of the coherent state in the early (late) mode,  $\mu_{\rm max}:=\mu_{\rm in}\eta_{S}$ represents the maximum possible intensity in each of the modes $e$ and $l$, $\phi\in(-\pi,\pi]$ refers to the relative phase between the modes $e$ and $l$, $\phi_e$ ($\phi_l$) is the phase of the coherent state in the mode $e$ ($l$), $\theta\in[0,\pi]$ represents the polar angle of the time-bin state, and
\begin{equation}\label{eq:phase_funct}
    \varphi_{\Delta} = \begin{cases} 0 & \text { if } \abs{\Delta}\leq\pi, \\
\pi &  \text { if } \pi<\abs{\Delta}\leq2\pi.\end{cases}
\end{equation}
Note that, with the definitions provided in~\cref{eq:transformations1}, one can rewrite the state of the joint time-bin system $el$ as
\begin{equation}\label{eq:state1rephrased}
    \ket{\sqrt{\mu}\cos(\theta/2)e^{i\phi_e}}_e \otimes \ket*{\sqrt{\mu}\sin(\theta/2)e^{i(\phi+\phi_e)}}_l.
\end{equation}

The role of Alice's classical measurements in the CPM is to determine the quantities $\theta$, $\phi$ and $\mu$ defined in~\cref{eq:transformations1}. The parameters $\mu$ and $\theta$ can be directly obtained from the intensities $\mu_e$ and $\mu_l$, which are simply determined by measuring the classical intensity of the early and late time slots. On the other hand, the absolute value of the relative phase $\phi$ can be computed as
\begin{equation}
    \abs{\phi}=\arccos\left(\frac{\mu_{(+)}-\mu_{(-)}}{4\sqrt{\mu_e\mu_l}}\right),
\end{equation}
where $\mu_{(\pm)}=\mu_e+\mu_l\pm2\sqrt{\mu_e\mu_l}\cos(\phi)$ are the classical intensities of the output pulses in a Faraday-Michelson interferometer (see~\cite{lu2023experimental} for further details). 

In particular, the bit/basis post-selection is carried out by the CPM based on the following predefined post-selection regions of the $(\theta,\phi)$-space (see~\cref{fig:scheme}b):
\begin{itemize}[leftmargin=15mm]
    \item[$(0_Z)$:] If $\theta<\Delta\theta_Z$, then the CPM outputs $\beta=Z$ and $a=0$;
    \item[$(1_Z)$:] If $\theta>\pi-\Delta\theta_Z$, then the CPM outputs $\beta=Z$ and $a=1$;
    \item[$(0_X)$:] If $\theta\in(\frac{\pi}{2}-\Delta\theta_X,\frac{\pi}{2}+\Delta\theta_X)$ and $\abs{\phi}<\Delta\phi_X$, the CPM outputs $\beta=X$ and $a=0$;
    \item[$(1_X)$:] If $\theta\in(\frac{\pi}{2}-\Delta\theta_X,\frac{\pi}{2}+\Delta\theta_X)$ and $\abs{\phi-\pi}<\Delta\phi_X$, the CPM outputs $\beta=X$ and $a=1$;
\end{itemize}
where the values of $\Delta\theta_Z$, $\Delta\theta_X$ and $\Delta\phi_X$ are fixed in advance. Note that, due to the symmetry of the post-selection regions, we can assume without loss of generality that the CPM indeed determines the value of $\phi$ ---instead of $\abs{\phi}$--- for each round. 

Under the condition that the basis and bit values have been successfully determined ---\textit{i.e.}, if the pair ($\theta$,$\phi$) lies in one of the prefixed post-selection regions--- the CPM determines also the intensity. Here we consider three post-selection intervals, each of them associated with a particular intensity $I\in\bar{I}\equiv\set{I_0,I_1,I_2}$. These are defined as
\begin{itemize}[leftmargin=15mm]
    \item[$(I_{0})$:] If $t_1\mu_{\rm max}\leq\mu<2\mu_{\rm max}$, then $I=I_{0}$;
    \item[$(I_{1})$:] If $t_2\mu_{\rm max}\leq\mu<t_1\mu_{\rm max}$, then $I= I_{1}$;
    \item[$(I_{2})$:] If $\mu<t_2\mu_{\rm max}$, then $I= I_{2}$;
\end{itemize}
where the two thresholds $t_1$ and $t_2$ are prefixed. Note that the previous definition of intervals could be modified to allow for any given round to contribute to more than one intensity set~\cite{lu2023experimental}. However, we do not consider this scenario here for simplicity.

Finally, we denote by $\Omega_{a,\beta}^{I}$ the post-selection region of the $(\theta,\phi,\mu)$-space associated with the bit value $a$, the basis $\beta$, and the intensity $I$. Besides, we conveniently define $\Omega_{\beta}^{I}:=\Omega_{0,\beta}^{I}\cup\Omega_{1,\beta}^{I}$, and denote as $p_{\Omega}$ the probability that Alice post-selects a particular region $\Omega$.

%%%%%%%%%%%%
\subsection{Information leakage}
In real-world scenarios, the extinction ratio of an IM is always finite, which means that the output state of the transmitter in each round is actually given by
\begin{equation}\label{eq:state1}
    %\ket*{\sqrt{\mu_{\rm max}}\left(e^{i\phi_1}+e^{i\phi_2}\right)/2}_e \otimes \ket{\sqrt{\mu_{\rm max}}\left(e^{i\phi_3}+e^{i\phi_4}\right)/2}_l\otimes\ket*{\lambda_{\phi_1^4}^{\omega}}_L =
    \ket*{\sqrt{\mu}\cos(\theta/2)e^{i\phi_e}}_e \otimes \ket*{\sqrt{\mu}\sin(\theta/2)e^{i(\phi+\phi_e)}}_l \otimes \ket*{\lambda_{\phi_1^4}^{\omega}}_L,
\end{equation}
where the label $L$ refers to the odd optical modes that leak information about the random phases $\phi_1,\cdots,\phi_4$. Namely, $L$ tags the odd time slots 1, 3 and 5, where the fifth mode refers to the first time bin of the subsequent round. That is, part of the sensitive information of a round is leaked in a time slot associated with its subsequent round. Specifically, the state of the leakage system $L$ can be written as
\begin{equation}\label{eq:leakageState1}
    \ket*{\lambda_{\phi_1^4}^{\omega}}_L=\ket*{\sqrt{\omega}/2\left(e^{i\phi_0}+e^{i\phi_1}\right)}_1
    \otimes \ket{\sqrt{\omega}/2\left(e^{i\phi_2}+e^{i\phi_3}\right)}_3 
    \otimes \ket{\sqrt{\omega}/2\left(e^{i\phi_4}+e^{i\phi_5}\right)}_5,
\end{equation}
where $\omega:=\mu_{\rm in}\eta_{S}\eta_{\rm IM}=\mu_{\rm max}\eta_{\rm IM}$, the parameter $\eta_{\rm IM}\ll 1$ denotes the transmittance of the IM ---when active--- and $\phi_0$ ($\phi_5$) represents the optical phase of the last (first) pulse of the previous (subsequent) pulse train. That is, if we use the superscript $u$ to indicate the round, then $\phi_0^u\equiv\phi_4^{u-1}$ and $\phi_5^u\equiv\phi_1^{u+1}$. Importantly, knowledge about $\phi_0$ and $\phi_5$ enhances Eve's ability to guess the settings of the present round. This means that, in principle, to account for all the information leakage one needs to consider all the transmitted rounds together. 

To simplify the analysis, however, we shall consider a more advantageous scenario for Eve that eliminates the need to evaluate all rounds. In this alternative fictitious scenario, Alice transmits not one but two optical systems in each of the odd time slots, such that Eve can always combine these two systems to reconstruct the original state prepared by Alice in the actual scenario. To be precise, we construct this alternative setting from the actual one by making the following substitutions in~\cref{eq:leakageState1},
\begin{equation}
\begin{split}
    \ket{\sqrt{\omega}/2\left(e^{i\phi_0}+e^{i\phi_1}\right)}_1
    \to\ket*{\sqrt{\omega/2}e^{i\phi_0}}_{1'}\otimes\ket*{\sqrt{\omega/2}e^{i\phi_1}}_{1},\\
    \ket{\sqrt{\omega}/2\left(e^{i\phi_4}+e^{i\phi_5}\right)}_5
    \to\ket*{\sqrt{\omega/2}e^{i\phi_4}}_{5}\otimes\ket*{\sqrt{\omega/2}e^{i\phi_5}}_{5'}.
\end{split}
\end{equation}
Note that Eve could always combine systems $1'$ and 1 (5 and $5'$) in a 50:50 beamsplitter to obtain the original state in system 1 (5) ---plus another extra system that she could simply discard---. Therefore, the secret-key rate in this conservative setting provides a valid lower bound of the actual secret-key rate. We apply this type of substitution to all the inter-round odd time slots. In doing so, we find that the state of the leakage system corresponding to a particular round can be assumed to be
\begin{equation}\label{eq:leakageState2}
    \ket*{\lambda_{\phi_1^4}^{\omega}}_L=
    \ket*{\sqrt{\omega/2}e^{i\phi_1}}_{1}
    \otimes \ket{\sqrt{\omega}/2\left(e^{i\phi_2}+e^{i\phi_3}\right)}_3 
    \otimes \ket*{\sqrt{\omega/2}e^{i\phi_4}}_{5},
\end{equation}
where $L\equiv 135$, and we have omitted systems $1'$ and $5'$ because they do not contain any information about the phases involved in the present round. In fact, we could also apply this type of substitution to all the intra-round time slots (\textit{i.e.}, system 3) to further simplify the analysis ---at the cost of introducing an extra leakage mode--- leading to the alternative definition $\ket*{\lambda_{\phi_1^4}^{\omega}}_L=
\bigotimes_{s=1}^4\ket*{\sqrt{\omega/2}e^{i\phi_s}}_{L_s}.$ Nonetheless, from now on we shall only consider~\cref{eq:leakageState2}.

%%%%%%%%%%%
\subsection{Post-selected density matrices}
Next we rewrite the state $\ket*{\lambda_{\phi_1^4}^{\omega}}_L$ introduced in~\cref{eq:leakageState2} as a function of the target variables $(\theta,\phi,\mu)$ to compute the mixed states emitted by Alice in the fictitious scenario. Note that the post-selection process is identical to that in~\cite{lu2023experimental}, as the probability density function $f(\theta,\phi,\mu)$ of the classical outcome $(\theta,\phi,\mu)$ is not affected by the information leakage. As explained in~\cref{appendix:transformations}, conditioned on a particular value of the target variables $(\theta,\phi,\mu)$, the original random phases $(\phi_1,\dots,\phi_4)$ take the values
\begin{equation}
\begin{aligned}\label{eq:substitutions}
    \phi_1&=\phi_e+s_e\frac{1}{2}\arccos{\left(\frac{2\mu\cos^2(\theta/2)}{\mu_{\rm max}}-1\right)},
    &\phi_3&=\phi_e+\phi+s_l\frac{1}{2}\arccos{\left(\frac{2\mu\sin^2(\theta/2)}{\mu_{\rm max}}-1\right)},\\
    \phi_2&=\phi_e-s_e\frac{1}{2}\arccos{\left(\frac{2\mu\cos^2(\theta/2)}{\mu_{\rm max}}-1\right)},
    &\phi_4&=\phi_e+\phi-s_l\frac{1}{2}\arccos{\left(\frac{2\mu\sin^2(\theta/2)}{\mu_{\rm max}}-1\right)},
\end{aligned}  
\end{equation}
where $s_e,s_l\in\set{1,-1}$ and all four $(s_e,s_l)$-transformations are equally likely. This means that, given the outcome $(\theta,\phi,\mu)$, the density operator $\sigma_{\theta,\phi,\mu}^{\omega}$ that describes the full transmitted system can be written as a classical mixture of the form
\begin{equation}\label{eq:averageCases}
    \sigma_{\theta,\phi,\mu}^{\omega}=\frac{1}{4}\sum_{s_e,s_l} \sigma_{\theta,\phi,\mu}^{\omega,(s_e,s_l)},
\end{equation}
where
\begin{equation}\label{eq:sigma_integral_phie}
\begin{split}
    %\sigma_{\theta,\phi,\mu}^{\omega,(s_e,s_l)}&=\frac{1}{2\pi}\int_0^{2\pi} \ketbra*{\sqrt{\mu}\cos(\theta/2)e^{i\phi_e}}{\cdot}_e
    %\otimes\ketbra*{\sqrt{\mu}\sin(\theta/2)e^{i(\phi+\phi_e)}}{\cdot}_l
    %\otimes\ketbra*{\lambda_{\theta,\phi,\mu,\phi_e}^{\omega,(s_e,s_l)}}{\cdot}_L d\phi_e,
    \sigma_{\theta,\phi,\mu}^{\omega,(s_e,s_l)}&=\frac{1}{2\pi}\int_0^{2\pi} \hat{\text{P}}_{\ket*{\sqrt{\mu}\cos(\theta/2)e^{i\phi_e}}_e}
    \otimes\hat{\text{P}}_{\ket*{\sqrt{\mu}\sin(\theta/2)e^{i(\phi+\phi_e)}}_l}
    \otimes\hat{\text{P}}_{\ket*{\lambda_{\theta,\phi,\mu,\phi_e}^{\omega,(s_e,s_l)}}_L} d\phi_e.
\end{split}
\end{equation}
Here, we use the notation $\hat{\text{P}}_{\ket{\psi}}\equiv\ketbra{\psi}{\psi}$, and $\ket*{\lambda_{\theta,\phi,\mu,\phi_e}^{\omega,(s_e,s_l)}}$ is obtained by applying the corresponding $(s_e,s_l)$-transformation from~\cref{eq:substitutions} to the state given in~\cref{eq:leakageState2}. Finally, upon post-selection of the region $\Omega_{a,\beta}^{I}$, the resulting mixture is the average over all conditional states $\sigma_{\theta,\phi,\mu}^{\omega}$ in this region. Namely,
\begin{equation}\label{eq:sigmaOmega1}
    \rho_{\abit,\abasis,I}^{\omega}=\frac{\langle \sigma_{\theta,\phi,\mu}^{\omega}\rangle_{\Omega_{a,\beta}^{I}}}{\langle 1\rangle_{\Omega_{a,\beta}^{I}}}=\frac{1}{\langle 1\rangle_{\Omega_{a,\beta}^{I}}}\iiint_{\Omega_{a,\beta}^{I}} f(\phi,\theta,\mu) \sigma_{\theta,\phi,\mu}^{\omega} d\theta d\phi d\mu,
\end{equation}
where $\langle 1\rangle_{\Omega}=\iiint_{\Omega} f(\phi,\theta,\mu) d\theta d\phi d\mu$, and~\cite{lu2023experimental,zapatero2023fully}
\begin{equation}
    f(\theta,\phi,\mu)=f_{\phi}(\phi)\frac{1}{\mu_{\rm max}\pi^2\sqrt{1-\frac{\mu}{\mu_{\rm max}}\cos^2(\theta/2)}\sqrt{1-\frac{\mu}{\mu_{\rm max}}\sin^2(\theta/2)}},
\end{equation}
with $f_{\phi}(\phi)=\frac{1}{2\pi}$.

In the absence of information leakage ---\textit{i.e.,} when $\omega=0$--- all four $(s_e,s_l)$-transformations result in the same transmitted state, and so one only needs to compute the average over $\phi_e$ in~\cref{eq:sigma_integral_phie}, which yields
\begin{equation}\label{eq:sigma_leak_ideal}
\begin{split}
   \sigma_{\theta,\phi,\mu}^{\omega=0}=&\frac{1}{2\pi}\int_0^{2\pi}e^{-\mu}\sum_{n_e,m_e}\sum_{n_l,m_l}\frac{(\sqrt{\mu}\cos(\theta/2))^{n_e+m_e}e^{i(n_e-m_e)\phi_e}}{\sqrt{n_e!m_e!}}\\
   &\times\frac{(\sqrt{\mu}\sin(\theta/2))^{n_l+m_l}e^{i(n_l-m_l)\phi}e^{i(n_l-m_l)\phi_e}}{\sqrt{n_l!m_l!}}\ketbra{n_e,n_l}{m_e,m_l}_{el} d\phi_e\\
   =&\sum_{n}\bar{\sigma}_{\theta,\phi,\mu}^{\omega=0,n}, 
\end{split}
\end{equation}
where we defined the sub-normalized ---\textit{i.e.}, with trace less than or equal to one--- density operators
\begin{equation}\label{eq:sigmaOmega0n}
\begin{split}
   \bar{\sigma}_{\theta,\phi,\mu}^{\omega=0,n}
   &=e^{-\mu}\sum_{n_e,m_e\leq n}\frac{(\sqrt{\mu}\cos(\theta/2))^{n_e+m_e}}{\sqrt{n_e!m_e!}}
   \frac{(\sqrt{\mu}\sin(\theta/2))^{2n-n_e-m_e}e^{-i(n_e-m_e)\phi}}{\sqrt{(n-n_e)!(n-m_e)!}}\ketbra{n_e,n-n_e}{m_e,n-m_e}_{el}.
\end{split}
\end{equation}
That is, after averaging over $\phi_e$, the resulting state $\sigma_{\theta,\phi,\mu}^{\omega=0}$ is restricted to a subspace of $\mathcal{P}_{\circ}\left(\mathcal{H}_e\otimes\mathcal{H}_l\right)$ ---the space of trace-one positive semi-definite operators acting on $\mathcal{H}_e\otimes\mathcal{H}_l$, with $\mathcal{H}_x$ being the Fock space in the mode $x$--- that consists of a direct sum of $n$-photon subspaces  where the $\bar{\sigma}_{\theta,\phi,\mu}^{\omega=0,n}$ live. For instance, for $n=1$ one obtains the pure state $\cos(\theta/2)\ket{1,0}_{el}+e^{i\phi}\sin(\theta/2)\ket{0,1}_{el}$, which lies in the single-photon qubit subspace $\mathcal{H}^{1}=\text{span}\set{\ket{0,1}_{el},\ket{1,0}_{el}}$. Since each of the $n$-photon subspaces has a finite number of basis states, one could in principle compute any of the postselected states given in~\cref{eq:sigmaOmega0n} numerically.

If $\omega> 0$, the relevant Hilbert space becomes $\mathcal{H}_e\otimes\mathcal{H}_l\otimes\mathcal{H}_1\otimes\mathcal{H}_3\otimes\mathcal{H}_5$, and then even the single-photon component ($n=1$) is described by a non-qubit subspace. In particular, now we have that
\begin{equation}\label{eq:sigmaLeakage1}
\begin{split}
   \sigma_{\theta,\phi,\mu}^{\omega,(s_e,s_l)}=&e^{-\mu-\mu_L(\theta,\phi,\mu)}\sum_{n}\sum_{\substack{\abs{\bf n}= n\\\abs{\bf m}= n}}
   \frac{(\sqrt{\mu}\cos(\theta/2))^{n_e+m_e}}{\sqrt{n_e!m_e!}}
   \frac{(\sqrt{\mu}\sin(\theta/2))^{n_l+m_l}e^{i(n_l-m_l)\phi}}{\sqrt{n_l!m_l!}}\\
   &\times\frac{\sqrt{\omega/2}^{n_1+m_1}e^{i(n_1-m_1)C(\theta,\mu)}}{\sqrt{n_1!m_1!}}
   \frac{\sqrt{\omega/2}^{n_5+m_5}e^{i(n_5-m_5)(\phi-S(\theta,\mu))}}{\sqrt{n_5!m_5!}}\\
   &\times\frac{r(\theta,\phi,\mu)^{n_3+m_3}e^{i(n_3-m_3)h(\theta,\phi,\mu)}}{\sqrt{n_3!m_3!}}
   \ketbra{n_e,n_l,n_1,n_3,n_5}{m_e,m_l,m_1,m_3,m_5}_{el135} \\
   =&\sum_{n}\Bar{\sigma}_{\theta,\phi,\mu}^{\omega,(s_e,s_l),n},
\end{split}
\end{equation}
where
\begin{align}\label{eq:fucntions_def}\small
    &C(\theta,\mu)=\frac{s_e}{2}\arccos\left(\frac{2\mu\cos^2(\theta/2)}{\mu_{\rm max}}-1\right),
    & &S(\theta,\mu)=\frac{s_l}{2}\arccos\left(\frac{2\mu\sin^2(\theta/2)}{\mu_{\rm max}}-1\right),\nonumber\\
    &r(\theta,\phi,\mu)=\sqrt{\frac{\omega(1+\cos\left(\phi+C(\theta,\mu)+S(\theta,\mu)\right)}{2}},
    & &h(\theta,\phi,\mu)=\arg(e^{-i C(\theta,\mu)}+e^{i(\phi + S(\theta,\mu))}),\\
    &\mu_L(\theta,\phi,\mu)=\omega+r^2(\theta,\phi,\mu),\nonumber
\end{align}
and $\textbf{n}\equiv n_e,n_l,n_1,n_3,n_5$, with $\abs{\textbf{n}}=n$ denoting the constraint $n_e+\dots+n_5=n$ (and similarly for $\textbf{m}$). Note that in~\cref{eq:fucntions_def} we have omitted the dependence of the functions on $s_e$ and $s_l$ for simplicity of notation. With this, we can define the sub-normalized $n$-photon state as
\begin{equation}
    \bar\sigma_{\theta,\phi,\mu}^{\omega,n}=\frac{1}{4}\sum_{s_e,s_l} \bar\sigma_{\theta,\phi,\mu}^{\omega,(s_e,s_l),n},
\end{equation}
which now lives in a ${n+4\choose n}$-dimensional Hilbert space. Combining this with~\cref{eq:sigmaOmega1}, one obtains, for the $\Omega_{a,\beta}^{I}$ region, that
\begin{equation}
    \rho_{\abit,\abasis,I}^{\omega}
    =\frac{1}{\langle 1\rangle_{\Omega_{a,\beta}^{I}}}\sum_{n=0}^{\infty}\langle \bar\sigma_{\theta,\phi,\mu}^{\omega,n}\rangle_{\Omega_{a,\beta}^{I}}
    =\sum_{n=0}^{\infty}p_{n|\Omega_{a,\beta}^{I},\omega}\rho_{\abit,\abasis,I}^{\omega,n},
\end{equation}
where 
\begin{equation}
p_{n|\Omega}^{\omega}=\frac{\Tr{\langle \bar\sigma_{\theta,\phi,\mu}^{\omega,n}\rangle_{\Omega}}}{\langle1\rangle_{\Omega}}
=\frac{\Big\langle\Tr{ \bar\sigma_{\theta,\phi,\mu}^{\omega,n}}\Big\rangle_{\Omega}}{\langle1\rangle_{\Omega}}
=\frac{\langle e^{-\mu-\mu_L(\theta,\phi,\mu)}[\mu+\mu_L(\theta,\phi,\mu)]^n/n!\rangle_{\Omega}}{\langle1\rangle_{\Omega}},
\end{equation}
and
\begin{equation}\label{eq:sigmaOmega_n}
    \rho_{\abit,\abasis,I}^{\omega,n}=\frac{\langle \bar\sigma_{\theta,\phi,\mu}^{\omega,n}\rangle_{\Omega_{a,\beta}^{I}}}{p_{n|\Omega_{a,\beta}^{I}}^{\omega}\langle1\rangle_{\Omega_{a,\beta}^{I}}}.
\end{equation}
\subsection{Security}\label{sec:security}
To prove the security of the protocol, we will adopt the traditional approach of proving the security of an equivalent entanglement-based scheme. As mentioned in~\cref{sec:PassiveTransmitter}, in each round of the actual protocol Alice's CPM performs some classical measurements to determine which mixed state $\rho_{\abit,\abasis,\aint}^{\omega}$ is being transmitted. 
Besides, we have that these transmitted states can be written as a classical mixture of $n$-photon states $\rho_{\abit,\abasis,\aint}^{\omega,n}$. 
This means, in particular, that one could consider an equivalent scenario in which, in the $n$-photon $\Omega_{Z}^{\aint}$-rounds, Alice prepares a purification $\ket*{\psi_{Z,\aint}^{\omega,n}}_{\SAnc\SShi\STra}$ of the state $\rho_{Z,I}^{\omega,n}=\frac{1}{2}\rho_{0,Z,\aint}^{\omega,n}+\frac{1}{2}\rho_{1,Z,\aint}^{\omega,n}$ with probability $p_{n|\Omega_{Z}^{\aint},\omega}$, where $\SAnc$ is an ancillary system in Alice's hands storing the encoded bit value, $\SShi$ is a purifying shield system, and $\STra\equiv el135$ is the joint transmitted system. In particular, this state is given by
\begin{equation}\label{eq:virtual_state}
    \ket*{\psi_{\abasis,I}^{\omega,n}}_{\SAnc\SShi\STra}=\frac{1}{\sqrt{2}}\left[\ket{0_{\beta}}_{\SAnc}\ket*{\rho_{0,\abasis,I}^{\omega,n}}_{\SShi\STra}+\ket{1_{\beta}}_{\SAnc}\ket*{\rho_{1,\abasis,I}^{\omega,n}}_{\SShi\STra}\right],
\end{equation} 
where $\ket*{\rho_{\abit,\abasis,I}^{\omega,n}}_{\SShi\STra}$ is a purification of $\rho_{\abit,\abasis,I}^{\omega,n}$. 

As described in~\cref{sec:PassiveTransmitter}, we assume that Alice and Bob generate key from the subset of $\Omega_{Z}^{I_0}$-rounds in which Bob obtains a detection in the $Z$ basis, and we denote them as the key-generation rounds. In addition, we consider the standard basis-independent detection efficiency condition at Bob's side. That is, we consider that the detection probability in his measurement apparatus does not depend on the measurement basis selected. In such situation, the detection losses can be decoupled from Bob's receiver and assumed to be in Eve's hands as part of the quantum channel. Consequently, one could consider the scenario in which Bob performs a quantum nondemolition measurement in each round to determine whether or not the incoming signal contains at least one photon ---so it will be detected with certainty--- followed by a $Z$ or $X$-basis measurement implemented with ideal detectors. In this setting, Eve's information on Alice's sifted key can be upper bounded from the single-photon phase-error rate $e_{\rm ph}$, defined as the ratio of bit errors that Alice and Bob would find within the single-photon key rounds if they measured their respective systems $A$ and $B$ ---with $B$ being the system that Bob receives after Eve's intervention on the transmitted system $T$--- in the $X$ basis. In particular, we have that the asymptotic secret-key rate of the protocol satisfies~\cite{koashi2009simple} 
\begin{equation}\label{eq:rate}
    R\geq p_{Z_B}p_{\Omega^{I_0}_{Z}}p_{1|\Omega^{I_0}_{Z}}^{\omega}Y_{Z,I_0}^{\omega,1,{\rm L}}\left[1-h_2\left(e_{\rm ph}^{\omega,{\rm U}}\right)\right]- p_{Z_B}p_{\Omega^{I_0}_{Z}} Q_{Z,I_0} f_{\rm EC} h_2(E_{Z,I_0}),
\end{equation}
where $Y_{Z,I_0}^{\omega,1,{\rm L}}$ is a lower bound on the single-photon yield $Y_{Z,I_0}^{\omega,1}$, $e_{\rm ph}^{\omega,{\rm U}}$ is an upper bound on the single-photon phase-error rate $e_{\rm ph}^{\omega}$, $Q_{\abasis,I}$ ($E_{\abasis,I}$) is the gain (bit-error rate) associated to the rounds in which Alice post-selects $\Omega_{\abasis}^{I}$ and Bob selects the measurement basis $\beta$, $f_{\rm EC}$ is the efficiency of the error correction protocol, and $h_2(p)=-p \log_2(p)-(1-p)\log_2(1-p)$ is the binary entropy function. Precise definitions of some of the previous quantities will be provided in subsequent subsections. Note that the lower bound on the secret-key rate provided by~\cref{eq:rate} assumes the typical scenario in which Alice and Bob distill key only from single-photon key rounds, therefore disregarding any amount of key material that they could distill from rounds in which $n\neq1$. Now we show how to calculate the quantities $Y_{Z,I_0}^{1,\rm L}$ and $e_{\rm ph}^{\rm U}$.

\subsubsection{Single-photon yield}\label{subsec:single_photon_yield}
Let us define the yield $Y_{\abasis,\aint}^{\omega,n}$ as the conditional probability that Bob observes a detection \textit{click} in his measurement apparatus given that Alice prepares $\rho_{\abasis,\aint}^{\omega,n}$ and he selects the basis $\abasis$. Unlike the case of an ideal decoy-state protocol without information leakage, here these yields depend on the intensity setting $\aint$, which prevents us from using the standard analysis to compute $Y_{Z,I_0}^{\omega,1,{\rm L}}$. Fortunately, this can be solved by invoking the quantum-coin argument~\cite{gottesman,Lo-Preskill,Lo-Ma,curras2023security} (see~\cref{appendix:quantum-coin-argument-Yields}), which allows us to impose some constraints between the yields associated with different intensity settings and construct a linear program to lower bound $Y_{Z,I_0}^{\omega,1}$.
Crucially, this method requires to compute lower bounds on the fidelities between photon-number states that differ on the post-selected intensity. To be precise, defining
\begin{equation}
    F_{I,J}^{\abasis,n}:=F(\rho_{\abasis,I}^{\omega,n},\rho_{\abasis,J}^{\omega,n}),
\end{equation}
where we omitted the dependency on $\omega$ for simplicity of notation and $F(\rho,\sigma):=\text{Tr}\set{\sqrt{\rho\sqrt{\sigma}\rho}}^2$, the linear program reads as
\begin{gather}\label{eq:LPYields}
\begin{aligned} 
    \textup{min}\quad& Y_{\abasis,I_0}^{\omega,1} \\
    \textup{s.t.}\quad 
    & \sum_{n=0}^{n_{\rm cut}}p_{n|\Omega_{\abasis}^{I}}^{\omega}Y_{\abasis,I}^{\omega,n} 
    \leq Q_{\abasis,I}\leq 
    1- \sum_{n=0}^{n_{\rm cut}}p_{n|\Omega_{\abasis}^{I}}^{\omega}(1-Y_{\abasis,I}^{\omega,n})\qquad (I \in\bar{I}),\\
    & \text{LCS}_{F_{I,J}^{\abasis,n}}^{\rm L}\left(Y_{\abasis,I}^{\omega,n}\right) \leq Y_{\abasis,J}^{\omega,n} \leq
    \text{LCS}_{F_{I,J}^{\abasis,n}}^{\rm U}\left(Y_{\abasis,I}^{\omega,n}\right),
    \qquad (n\leq n_{\rm cut}, I\neq J\in\bar{I}),\\
    & 0\leq Y_{\abasis,I}^{\omega,n} \leq 1
    \qquad (n\leq n_{\rm cut}, I\in\{I_0,I_1,I_2 \}),
\end{aligned}
\end{gather}
where $n_{\rm cut}$ is the threshold photon-number considered in the linear program, $\text{LCS}^{K}_{z}(y)=G_{z}^{K}(\tilde{y}) + (y-\tilde{y})G_{z}^{K '}(\tilde{y})$ with $K\in\set{\rm U,L}$,
\begin{equation}\label{eq:Gfunction}
\begin{aligned}
G_{z}^{\rm L}\left(y\right) = \begin{cases} g_{-}(y, z) & \text { if } y>1-z \\
0 & \text { otherwise }\end{cases},
\quad\quad\quad\quad G_{z}^{\rm U}(y) = \begin{cases}g_{+}(y, z) & \text { if } y<z \\
1 & \text { otherwise }\end{cases},
\end{aligned}
\end{equation}
$g_{\pm}(y, z)=y+(1-z)(1-2 y) \pm 2 \sqrt{z(1-z) y(1-y)}$, $\tilde{y}$ is an arbitrary guess of $y$, and $G_{z}^{K '}$ is the derivative of $G_{z}^{K}$. 

The decoy-state constraints of the linear program in~\cref{eq:LPYields} are straightforwardly derived from the equality $Q_{\abasis,I}=\sum_{n=0}^{\infty}p_{n|\Omega_{\abasis}^{I}}^{\omega}Y_{\abasis,I}^{\omega,n}$, while the constraints included in the second line come, as already mentioned above, from the quantum-coin argument (further details can be found in \cref{appendix:quantum-coin-argument-Yields}). The quantities $\tilde{Y}_{\abasis,I}^{\omega,n}$ can be obtained, \textit{e.g}, from a model of the quantum channel, or they can be optimized numerically. Importantly, one can substitute the fidelities $F_{I,J}^{\abasis,n}$ required in the linear program by some lower bounds $F_{I,J}^{\abasis,n,\text{L}}$, and the result is still a valid lower bound of the single-photon yield $Y_{\abasis,I_0}^{\omega,1}$. We show how to calculate the bounds $F_{I,J}^{\abasis,n,\text{L}}$ in~\cref{appendix:EfficientFidelity}.

\subsubsection{Phase-error rate}
To upper bound the phase-error rate $e_{\rm ph}$ one must first estimate an upper bound on the single-photon bit-error rate in the $X$ basis $e_{X,I_0}^{\omega,1}$, which is defined as the probability of a bit error given that Alice prepares a single-photon state in the $X$ basis with intensity $I_0$, and Bob measures it also in the $X$ basis. For this, we follow a similar technique as for the yield estimation. Specifically, we construct a linear program that allows us to upper bound the quantities $\Gamma_{0,X,I_0}^{\omega,1}$ and $\Gamma_{1,X,I_0}^{\omega,1}$, where $\Gamma_{a,X,I}^{\omega,1}$ represents the bit-error probability given that Alice transmits the state $\rho^{\omega,1}_{\abit,X,I}$ and Bob selects the $X$ basis. The linear program is
\begin{gather}\label{eq:LPbiterror}
\begin{aligned} 
    \textup{max}\quad& \Gamma_{a,X,I_0}^{\omega,1} \\
    \textup{s.t.}\quad
    &\sum_{n=0}^{n_{\rm cut}}p_{n|\Omega_{\abit,X}^{I}}^{\omega}\Gamma_{\abit,X,I}^{\omega,n} 
    \leq E_{a,X,I}Q_{a,X,I}\leq 
    1- \sum_{n=0}^{n_{\rm cut}}p_{n|\Omega_{\abit,X}^{I}}^{\omega}(1-\Gamma_{\abit,X,I}^{\omega,n})\qquad (I\in\bar{I}),\\
    &\text{LCS}_{F_{I,J}^{a,X,n}}^{\rm L}\left(\Gamma_{\abit,X,I}^{\omega,n}\right) 
    \leq \Gamma_{\abit,X,J}^{\omega,n} \leq
    \text{LCS}_{F_{I,J}^{a,X,n}}^{\rm U}\left(\Gamma_{\abit,X,I}^{\omega,n}\right)
    \qquad (n\leq n_{\rm cut},  I\neq J\in\bar{I}),\\
    & 0\leq \Gamma_{\abit,X,I}^{\omega,n} \leq 1 
    \qquad (n\leq n_{\rm cut},  I\in\bar{I}),
\end{aligned}
\end{gather}
where $F_{I,J}^{a,\abasis,n}:=F(\rho_{\abit,\abasis,I}^{\omega,n},\rho_{\abit,\abasis,J}^{\omega,n})$, and $E_{a,X,I}$ ($Q_{a,X,I}$) denotes the observed bit-error rate (gain) of the rounds in which Alice postselects $\Omega_{a,X}^{I}$ and Bob selects the $X$ basis. With this, we can compute an upper bound on the bit-error probability of the postselected $\Omega_{X}^{I}$ single-photon states, namely,
\begin{equation}
    \Gamma_{X,I_0}^{\omega,1,{\rm U}}:=\frac{1}{2}\left(\Gamma_{0,X,I_0}^{\omega,1,{\rm U}}+\Gamma_{1,X,I_0}^{\omega,1,{\rm U}}\right),
\end{equation}
with $\Gamma_{a,X,I_0}^{\omega,1,{\rm U}}$ being the outcome of the linear program in~\cref{eq:LPbiterror}. Therefore, an upper bound on the single-photon $X$-basis bit-error rate can be directly computed as
\begin{equation}\label{eq:biterrorrateX}
    e_{X,I_0}^{\omega,1,{\rm U}}:=\frac{\Gamma_{X,I_0}^{\omega,1,{\rm U}}}{Y_{X,I_0}^{\omega,1,{\rm L}}},
\end{equation}
where the lower bound $Y_{X,I_0}^{\omega,1,{\rm L}}$ is obtained via~\cref{eq:LPYields}.

It only remains to link $e_{X,I_0}^{\omega,1,{\rm U}}$ to the phase-error rate $e_{\rm ph}^{\omega}$. For this, we invoke again the quantum-coin argument, which leads to the bound (see~\cref{appendix:quantum-coin-imbalance-ph})
\begin{equation}\label{eq:phase_error_bound1}
    e_{\rm ph}^{\omega}\leq e_{\rm ph}^{\omega,{\rm U}}:= G_{F_{ZX}'}^{\rm U}\left(e_{X,I_0}^{\omega,1,{\rm U}}\right),
\end{equation}
where the function $G_{z}^{\rm U}$ is defined in~\cref{eq:Gfunction}, and the quantity $F_{ZX}'$ is defined as
\begin{equation}
    F_{ZX}':=\left(1-\frac{1-\text{Re}\set{\braket*{\psi_{Z,I_0}^{\omega,1}}{\psi_{X,I_0}^{\omega,1}}}}{Y_{ZX\text{-coin}}^{\rm L}}\right)^2,
\end{equation}
with $Y_{ZX\text{-coin}}^{\rm L}=(Y_{Z,I_0}^{\omega,1,{\rm L}}+Y_{X,I_0}^{\omega,1,{\rm L}})/2$. We show how to lower bound $\text{Re}\set{\braket*{\psi_{Z,I_0}^{\omega,1}}{\psi_{X,I_0}^{\omega,1}}}$ in~\cref{appendix:quantum-coin-imbalance}.

The analysis we introduced in the section is versatile and can be readily adapted to a wide range of transmitters (see~\cref{sec:OILtransmitter}). Nevertheless, it does not account for some specific characteristics that are unique to a passive transmitter that uses post-selection. In~\cref{appendix:alternativeSec} we introduce a refined security analysis tailored for this type of passive transmitters that can further enhance the protocol's performance, particularly in scenarios where information leakage is very small.
\section{Modulator-free transmitter based on optical injection locking}\label{sec:OILtransmitter}
\begin{figure}
    \centering
    \includegraphics[width=0.50\columnwidth]{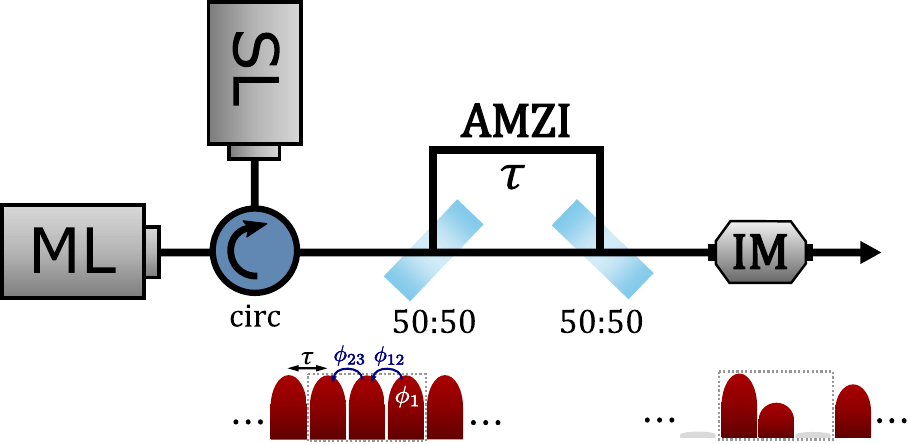}
    \caption{Schematic representation of the modulator-free transmitter proposed in~\cite{lo2023simplified}. Two lasers are placed in an optical injection locking configuration. Each round, the master laser (ML) is gain-switched with an electrical signal that introduces some controlled perturbations within the amplitude profile of each emitted pulse. This results in two consecutive controlled phase shifts $\phi_{12}$ and $\phi_{23}$ with respect the original random phase $\phi_{1}$ of the pulse. Each pulse emitted by the ML passes through a circulator (circ) to reach the slave laser (SL), which is gain-switched at a higher clock-rate such that it emits three shorter pulses per ML's pulse, each of them inheriting a different phase (see~\cite{lo2023simplified} for further details). The resulting pulses are directed to an asymmetric Mach-Zehnder interferometer (AMZI), followed by an intensity modulator (IM) that attenuates the first pulse of each three-pulses sequence to prevent information leakage.
    \label{fig:schemeLucamarini}}
\end{figure}

As already mentioned, the security analysis introduced in the previous section can be adapted as well to evaluate other modulator-free QKD transmitters, as, \textit{e.g.}, that recently presented in~\cite{lo2023simplified}. This is a time-bin decoy-state BB84 transmitter where the phase and intensity of the output signals are actively set by Alice without the need of intensity or phase modulators. The scheme is depicted in~\cref{fig:schemeLucamarini}. The transmitter utilizes optical injection locking between a master and a slave laser diode. The master laser is gain-switched to produce pulses with random phases that seed the slave laser, which also operates in gain-switching mode emitting three short pulses per master pulse. Carefully timed electrical perturbations in the master laser's driving signal induce specific phase shifts between slave pulses, enabling the precise control of their relative phases. The pulses are then interfered in an AMZI, resulting in two output pulses with finely tuned intensities and relative phases, and an additional pulse that is not used in the communication (see~\cite{lo2023simplified} for further details). That is, similar to the passive scheme examined earlier, this transmitter produces pulses that, while not serving as quantum carriers, still contain some sensitive information about Alice's quantum signals. Importantly, as already mentioned, even if Alice employs an IM to actively block these side-channel systems, the modulator's finite extinction ratio will inevitably result in some residual leakage that must be accommodated in the security proof. In~\cite{lo2023simplified}, the authors seem to provide an heuristic argument to justify that the unwanted pulses may not affect the protocol's security. Notwithstanding, our analysis indicates that sufficiently attenuating these pulses is essential to ensure the security of the decoy-state QKD scheme.

\subsection{Transmitted states}
This transmitter generates, in each round $u=1,\dots,N$, the following sequence of pulses~\cite{lo2023simplified}
\begin{equation}\label{eq:OILsequence1}
    \ket*{\sqrt{\mu_{\rm in}}e^{i\phi_1^u}}_{1_u}
    \ket*{\sqrt{\mu_{\rm in}}e^{i(\phi_1^u+\phi_{12}^u)}}_{2_u}
    \ket*{\sqrt{\mu_{\rm in}}e^{i(\phi_1^u+\phi_{12}^u+\phi_{23}^u)}}_{3_u},
\end{equation}
where the phase $\phi_1^{u}$ is uniformly random, and the relative phases $\phi_{12}^u$ and $\phi_{23}^u$ are directly controlled by Alice. The pulse train given by~\cref{eq:OILsequence1} then passes through an AMZI, resulting in the following state in the modes $1_u2_u3_u1_{u+1}$,
\begin{equation}\label{eq:OILsequence2}
\begin{split}
    \ket{\frac{\sqrt{\mu_{\rm in}}}{2}(e^{i\phi^{u-1}_P}+e^{i\phi_1^u})}_{1_u}
    \ket{\frac{\sqrt{\mu_{\rm in}}}{2}e^{i\phi_1^u}(1+e^{i\phi_{12}^u})}_{2_u}
    \ket{\frac{\sqrt{\mu_{\rm in}}}{2}e^{i(\phi_1^u+\phi_{12}^u)}(1+e^{i\phi_{23}^u})}_{3_u}
    \ket{\frac{\sqrt{\mu_{\rm in}}}{2}(e^{i(\phi_1^u+\phi_{12}^u+\phi_{23}^u)}+e^{i\phi_{F}^{u+1}})}_{1_{u+1}},
\end{split}
\end{equation}
where $\phi^{u-1}_P\equiv\phi_1^{u-1}+\phi_{12}^{u-1}+\phi_{23}^{u-1}$ and $\phi_{F}^{u+1}\equiv\phi^{u+1}_1$. Now, we rename the systems $2_u$ and $3_u$ as the early ($e$) and late ($l$) modes, respectively, as these are the modes in which Alice encodes the quantum information. Importantly, although systems $1_u$ and $1_{u+1}$ are not used as quantum carriers, they still contain information about Alice's transmitted states. Thus, here we consider that Alice places an IM after the AMZI, such that all these systems ---\textit{i.e.}, the systems $1_u$ from all rounds $u$--- are severely attenuated. Importantly, as in~\cref{sec:PassiveTransmitter}, the behavior of this IM is fixed and independent of Alice's setting choices. If we omit the round indices, and we rename systems $1_u$ and $1_{u+1}$ as $P$ and $F$, respectively, the state at the output of the IM can be written as
\begin{equation}\label{eq:OILsequence3}
\begin{split}
    \ket{\sqrt{\omega/2}(e^{i\phi_P}+e^{i\phi_1})}_{P}
    \ket{\frac{\sqrt{\mu_{\rm in}}}{2}e^{i\phi_1}(1+e^{i\phi_{12}})}_{e}
    \ket{\frac{\sqrt{\mu_{\rm in}}}{2}e^{i(\phi_1+\phi_{12})}(1+e^{i\phi_{23}})}_{l}
    \ket{\sqrt{\omega/2}(e^{i(\phi_1+\phi_{12}+\phi_{23})}+e^{i\phi_F})}_{F},
\end{split}
\end{equation}
where $\omega=\frac{\eta_{\rm IM}\mu_{\rm in}}{2}$, and $\eta_{\rm IM}$ is the transmittance of the IM used to block the undesired pulses.

As in~\cref{sec:PassiveTransmitter}, we simplify the security analysis by proving the security of an alternative fictitious scenario that is always more beneficial for Eve, such that a lower bound on the secret-key rate for such fictitious scenario also holds for the actual scenario. In particular, let us consider that, instead of directly transmitting the state described in~\cref{eq:OILsequence3}, Alice first substitutes each of the leakage systems $P$ and $F$ by two separate systems as follows
\begin{equation}
\begin{split}
    \ket*{\sqrt{\omega/2}(e^{i\phi_P}+e^{i\phi_1})}_{P}&\to 
    \ket*{\sqrt{\omega}e^{i\phi_P}}_{P'}\ket*{\sqrt{\omega}e^{i\phi_1}}_{P},\\
    \ket*{\sqrt{\omega/2}(e^{i(\phi_1+\phi_{12}+\phi_{23})}+e^{i\phi_F})}_{F}&\to 
    \ket*{\sqrt{\omega}e^{i\phi_1+\phi_{12}+\phi_{23}}}_{F}\ket*{\sqrt{\omega}e^{i\phi_F}}_{F'}.
\end{split}
\end{equation}
With this, now the transmitted state in a particular round can be simplified to
\begin{equation}\label{eq:OILstate1}
\begin{split}
    \ket{\frac{\sqrt{\mu_{\rm in}}}{2}e^{i\phi_1}(1+e^{i\phi_{12}})}_{e}
    \ket{\frac{\sqrt{\mu_{\rm in}}}{2}e^{i(\phi_1+\phi_{12})}(1+e^{i\phi_{23}})}_{l}
    \ket{\lambda}_{L},
\end{split}
\end{equation}
where 
\begin{equation}
    \ket*{\lambda}_{L}=\ket*{\sqrt{\omega}e^{i\phi_1}}_{P}\otimes\ket*{\sqrt{\omega}e^{i(\phi_1+\phi_{12}+\phi_{23})}}_{F},
\end{equation}
with $L\equiv PF$, is the state of the relevant leakage systems for the considered round, and systems $P'$ and $F'$ have been omitted because they play no role in the present round. 

If now, as in~\cref{sec:PassiveTransmitter}, we define the quantities $\mu_e=\mu_{\rm in}(1+\cos(\phi_{12}))/2$ and $\mu_l=\mu_{\rm in}(1+\cos(\phi_{23}))/2$, then \cref{eq:OILstate1} can be rewritten as
\begin{equation}\label{eq:OILstate2}
    \ket*{\sqrt{\mu_e}e^{i(\phi_1+\frac{\phi_{12}}{2})}}_{e}
    \ket*{\sqrt{\mu_l}e^{i(\phi_1+\phi_{12}+\frac{\phi_{23}}{2})}}_{l}
    \ket*{\sqrt{\omega}e^{i\phi_1}}_{P}
    \ket*{\sqrt{\omega}e^{i(\phi_1+\phi_{12}+\phi_{23})}}_{F}.
\end{equation}
Then, upon averaging on the uniform phase $\phi_1$, the previous state leads to the following mixture
\begin{equation}\small
\begin{split}\label{eq:sigma_OIL}
   \sigma_{\phi_{12},\phi_{23}}^{\omega}=&e^{-\mu-2\omega}\sum_{n}\sum_{\substack{\abs{\bf n}= n\\ \abs{\bf m}= n}}
   \frac{\sqrt{\mu_e}^{n_e+m_e}e^{i\frac{\phi_{12}}{2}(n_e-m_e)}}{\sqrt{n_e!m_e!}}
   \frac{\sqrt{\mu_l}^{n_l+m_l}e^{i(\phi_{12}+\frac{\phi_{23}}{2})(n_l-m_l)}}{\sqrt{n_l!m_l!}}\\
   &\times\frac{\sqrt{\omega}^{n_P+m_P+n_F+m_F}e^{i(\phi_{12}+\phi_{23})(n_F-m_F)}}{\sqrt{n_P!m_P!n_F!m_F!}}
   \ketbra{n_e,n_l,n_P,n_F}{m_e,m_l,m_L,m_F}_{elPF} \\
   =&\sum_{n}\Bar{\sigma}_{\phi_{12},\phi_{23}}^{\omega,n},
\end{split}
\end{equation}
where $\mu=\mu_e+\mu_l$, $\textbf{n}\equiv n_e,n_l,n_1,n_L,n_F$, $\abs{\textbf{n}}=n$ is a shorthand for $n_e+\dots+n_F=n$ (and similarly for $\textbf{m}$), and $\sum_{n}\Bar{\sigma}_{\phi_{12},\phi_{23}}^{\omega,n}$ is a sub-normalized state with $n$ photons distributed across the modes $elPF$.

From~\cref{eq:OILstate1}, it is clear that Alice can generate the desired BB84 states. In particular, she can generate $Y$-basis and $Z$-basis time-bin BB84 states. Nonetheless, in order to follow exactly the same notation as in~\cref{sec:PassiveTransmitter}, from now on we will rename the $Y$ basis (here used for key generation) as the $Z$ basis, and the $Z$ basis (here used for parameter estimation) as the $X$ basis. Besides, we denote by $\rho_{\abit,\abasis,I}^{\omega}$ the final states prepared by Alice when she selects the bit $a$, the basis $\beta$, and the intensity setting $I$, while $\ket*{\gamma_{\abit,\abasis,I}^{\omega,n}}$ denotes its corresponding $n$-photon contribution. In particular, these states can be computed from \cref{eq:sigma_OIL} given the following assignments~\cite{lo2023simplified}:
\begin{itemize}[leftmargin=15mm]
    \item[($Z$):] For the basis $\abasis=Z$, the bit $\abit=0$ ($\abit=1$) and the signal intensity setting $I=I_0$, Alice sets $\phi_{12}=\pi/2$ and $\phi_{23}=\pi/2$ ($\phi_{12}=-\pi/2$ and $\phi_{23}=-\pi/2$).
    \item[($X$):] For the basis $\abasis=X$ and the bit $\abit=0$ ($\abit=1$), Alice sets $\phi_{12}=\kappa\pi$ and $\phi_{23}=\pi$ ($\phi_{12}=\pi$ and $\phi_{23}=\kappa\pi$), for some $\kappa\in[0,1]$. Each value of $\kappa$ defines a different intensity. For example, $\kappa=0$ ($\kappa=1$) corresponds to the signal intensity $I_0$ setting (vacuum intensity).
\end{itemize}
Importantly, the sum $\phi_{12}+\phi_{23}$ is independent of the bit value $a$, and in fact it is also independent of the basis choice $\beta$ for the signal intensity setting $I_0$. In particular, this implies that, for the $I_0$ rounds, the leakage system $L$ only reveals information about the global phase $\phi_1$, which simplifies the security analysis. Also, note that, in principle, this transmitter allows to generate as many decoy states as desired, but only in the $X$ basis. For simplicity, below we will assume that Alice prepares perfect states ---apart from the information leakage--- but we remark that some imperfections in the state preparation process ---like, \textit{e.g.}, those produced by random fluctuations in the relative phases $\phi_{12}$ and $\phi_{23}$--- could be readily incorporated into the analysis by averaging the transmitted $n$-photon states, similar to our approach in~\cref{sec:PassiveTransmitter}.

\subsection{Security}
Let us consider that Alice uses three intensities ($I_0$, $I_1$, and $I_2$) in the test basis. As in~\cref{sec:security}, one can obtain a lower (upper) bound $Y_{X,I_0}^{\omega,1,{\rm L}}$ ($e_{X,I_0}^{\omega,1,{\rm U}}$) on the $X$-basis single photon yield (bit-error rate) associated with the intensity $I_0$ via~\cref{eq:LPYields,eq:LPbiterror,eq:biterrorrateX}, now computing the fidelities $F_{I,J}^{X,n}$ and $F_{I,J}^{a,X,n}$ from the states defined in this section. Nonetheless, as this transmitter does not generate decoy intensities in the $Z$ basis, we cannot use the decoy-state method to estimate the $Z$-basis single-photon yield, as we did in~\cref{sec:PassiveTransmitter}. Notably, however, the $I_0$ single-photon states emitted by this transmitter satisfy $\rho_{Z,I_0}^{\omega,1}=\rho_{X,I_0}^{\omega,1}$. This means that Eve cannot distinguish between these two kind of rounds, and so one can directly set $Y_{Z,I_0}^{\omega,1,{\rm L}}=Y_{X,I_0}^{\omega,1,{\rm L}}$. Nevertheless, we note that, even in a general setting where $\rho_{Z,I_0}^{\omega,1}\neq\rho_{X,I_0}^{\omega,1}$, one could still bound the desired quantities via, again, the quantum-coin argument. Specifically, we have that
\begin{equation}\label{eq:yield_ineq1}
    Y_{Z,I_0}^{\omega,1}\geq Y_{Z,I_0}^{\omega,1,{\rm L}}:= G_{F_{ZX}}^{\rm L}\left(Y_{X,I_0}^{\omega,1,{\rm L}}\right),
\end{equation}
where $F_{ZX}=F(\rho_{Z,I_0}^{\omega,n},\rho_{X,I_0}^{\omega,n})$, with $\rho_{\beta,I}^{\omega,n}:=\frac{1}{2}\ketbra*{\gamma_{0,\abasis,I}^{\omega,n}}{\gamma_{0,\abasis,I}^{\omega,n}}+\frac{1}{2}\ketbra*{\gamma_{1,\abasis,I}^{\omega,n}}{\gamma_{1,\abasis,I}^{\omega,n}}$. Of course, in the particular setting we are considering, we have that $F_{ZX}=1$, and the equality $Y_{Z,I_0}^{\omega,1,{\rm L}}=Y_{X,I_0}^{\omega,1,{\rm L}}$ is recovered from~\cref{eq:yield_ineq1}. Similarly, an upper bound on phase-error rate can be obtained from the bit-error rate in the $X$ basis as 
\begin{equation}
\begin{split}
    e_{\rm ph}^{\omega}&\leq e_{\rm ph}^{\omega,{\rm U}}:= G_{F_{ZX}'}^{\rm U}\left(e_{X,I_0}^{\omega,1,{\rm U}}\right), 
\end{split}
\end{equation}
where
\begin{equation}
\begin{split}
    F_{ZX}'&=\left(1-\frac{1-\text{Re}\set{\braket*{\psi_{Z,I_0}^{\omega,1}}{\psi_{X,I_0}^{\omega,1}}}}{Y_{ZX\text{-coin}}^{\rm L}}\right)^2,\\
    \ket*{\psi_{\abasis,I}^{\omega,1}}_{\SAnc\STra}&=\frac{1}{\sqrt{2}}\left[\ket{0_{\abasis}}_{\SAnc}\ket*{\gamma_{0,\abasis,I}^{\omega,1}}_{\STra}+\ket{1_{\abasis}}_{\SAnc}\ket*{\gamma_{1,\abasis,I}^{\omega,1}}_{\STra}\right],
\end{split}
\end{equation}
and $Y_{ZX\text{-coin}}^{\rm L}=\frac{1}{2}\left(Y_{Z,I_0}^{\omega,1,{\rm L}}+Y_{X,I_0}^{\omega,1,{\rm L}}\right)$. Finally, the secret-key rate in this scenario is given by
\begin{equation}
    R\geq p_{Z_AZ_B}p_{1|I_0,Z}Y_{Z,I_0}^{\omega,1,{\rm L}}\left[1-h_2(e_{\rm ph}^{\omega,{\rm U}})\right]- p_{Z_AZ_B} Q_{Z,I_0} f_{\rm EC} h_2(E_{Z,I_0}),
\end{equation}
where $p_{Z_AZ_B}$ is the probability that both Alice and Bob select the $Z$ basis, $p_{1|I_0,Z}$ is the probability that Alice emits a single-photon signal given that she selects the $Z$ basis, and all the remaining quantities are defined analogously to those in~\cref{eq:rate}.
\section{Simulations}\label{sec:Results}
In this section we investigate how the performance of the two previous QKD transmitters can be affected by the finite extinction ratio of practical IMs. For the simulations we use a typical decoy-state BB84 channel model, though the expected values of the observed quantities ---like the gain and the bit-error rate--- must be averaged over the corresponding post-selection regions in the case of the passive transmitter (see~\cref{appendix:chanel_model}). We also use this channel model to compute the reference yields and bit-error probabilities. For simplicity, we shall consider in both scenarios that Bob uses an active BB84 receiver with two identical detectors with dark-count probability $p_d=10^{-6}$ \cite{gobby2004quantum,ma2005practical} and perfect detection efficiency. Note, however, that accounting for a finite detection efficiency is straightforward as it only affects the channel model. Besides, we consider a typical fiber-loss coefficient $\alpha_{\rm dB}=0.2$ dB/km, and an error correction efficiency $f_{\rm EC}=1.16$.

\begin{figure}
    \centering
    \includegraphics[width=0.55\columnwidth]{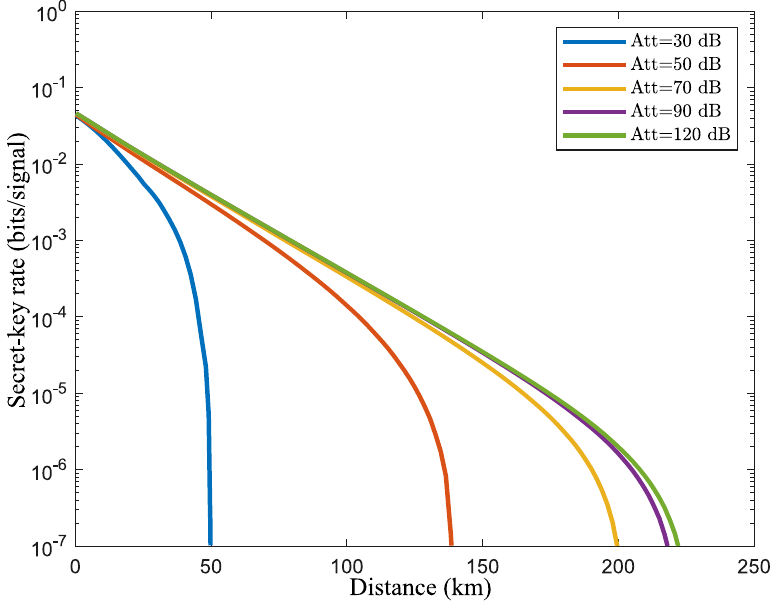}
    \caption{Asymptotic secret-key rate vs distance for the fully passive transmitter based on post-selection described in~\cref{sec:PassiveTransmitter} for different values of the attenuation Att (in dB) introduced by Alice's IM to block the undesired pulses. We consider a decoy-state BB84 protocol with three intensity settings. In the simulations, we numerically optimize over the parameters $\mu_{\rm max}$ and $\Delta\theta_Z$. For the simulations, we use the refined analysis introduced in~\cref{appendix:alternativeSec}. See~\cref{sec:Results} for further details.\label{fig:Passive_Tx_SKR}}
\end{figure}

\cref{fig:Passive_Tx_SKR} shows the asymptotic secret-key rate for the decoy-state BB84 protocol with the passive transmitter introduced in~\cref{sec:PassiveTransmitter} as a function of the channel distance and for different values of the overall attenuation $\text{Att}$ (in dB) provided by Alice's IM, which can be linked to its transmittance through $\eta_{\rm IM}=10^{-\frac{\text{Att}}{10}}$. Specifically, for the simulations we employ the refined analysis introduced in~\cref{appendix:alternativeSec} (we refer the reader to this appendix for a comparison with the general analysis introduced in~\cref{sec:security}). We see that the secret-key rate of the protocol is severely undermined when the attenuation provided by the IM is not sufficiently high. Nevertheless, the protocol's performance remains closely aligned with the ideal scenario (illustrated here by setting Att=120 dB, as no discernible difference is observed beyond this point) when $\text{Att}\gtrsim90$ dB. In the simulations, we consider that Alice post-selects three different intensity regions, as explained in~\cref{sec:PassiveTransmitter}. To define these regions, we fix for simplicity $t_1=0.05$ and $t_2=0.01$, and we optimize the value of $\mu_{\rm max}$ for each distance. We also optimize the value of $\Delta\theta_Z$, while we fix $\Delta\theta_X=0.11$ and $\Delta\phi_X=0.09$, which are the same values used in~\cite{lu2023experimental} but squeezed by a factor $\sim4$, as this leads to a better performance in the asymptotic-key regime.
Finally, we set in the linear programs $n_{\rm cut}=4$, and we use the approximations introduced in~\cref{appendix:EfficientFidelity} to speed up the optimization. In particular, for $n\leq 2$ we compute the full density matrices $\rho_{\abit,\abasis,I}^{\omega,n}$, while for $n=3$ and $n=4$ we disregard the matrix entries associated with states in which the leakage system contains more than one photon ---as explained in~\cref{appendix:EfficientFidelity}--- resulting in lower bounds on the fidelity that are only slightly looser than those obtainable when considering the full matrices.

\begin{figure}
    \centering
    \includegraphics[width=0.55\columnwidth]{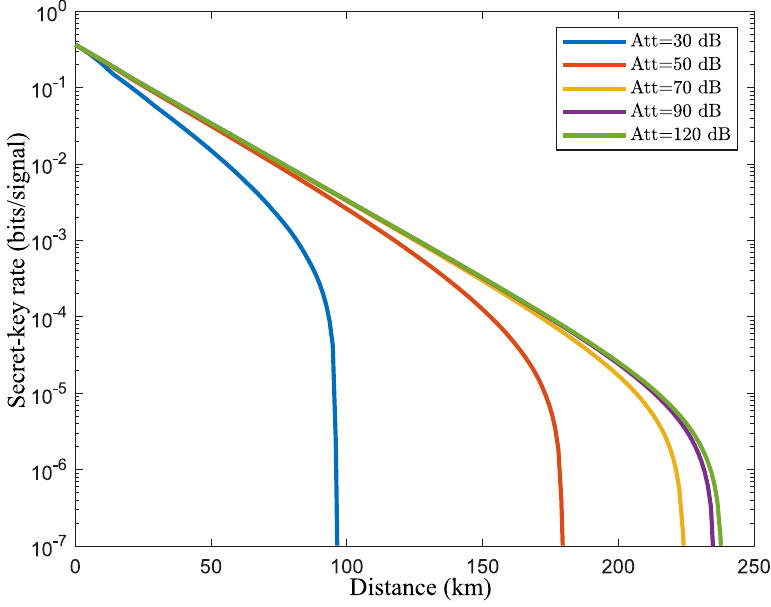}
    \caption{Asymptotic secret-key rate vs distance for the modulator-free transmitter proposed in~\cite{lo2023simplified} as a function of the attenuation Att (in dB) introduced by Alice's IM to block the undesired pulses. We consider a decoy-state BB84 protocol with three intensity settings in the test basis, and a single intensity in the key-generation basis. In the simulations, we numerically optimize the value of the two highest intensities. See~\cref{sec:Results} for further details.\label{fig:OIL_Tx_SKR}}
\end{figure}

In~\cref{fig:OIL_Tx_SKR} we show the results for the modulator-free transmitter introduced in~\cref{sec:OILtransmitter} based on optical injection locking. We consider a decoy-state BB84 protocol with three intensities in the $X$ basis, the two highest intensities being optimized for each distance, while we fix the weakest intensity to $I_2=10^{-4}$. Moreover, we fix for the simulations $n_{\rm cut}=4$. As expected, the secret-key rate obtained in this scenario outperforms that achieved with the fully passive transmitter introduced in~\cite{lu2023experimental}. 
The reason for this is twofold.  
Firstly, while the transmitter in~\cite{lo2023simplified} allows to prepare perfect BB84 states with any desired probability, the passive transmitter faces a trade-off, as small post-selection regions ---required to prepare close-to-perfect BB84 states--- inevitably come at the price of a large sifting penalty. In this sense, we note that a direct comparison between these two schemes is not entirely fair, as the transmitter in~\cite{lo2023simplified} is not passive ---as it requires active modulation of a laser diode to prepare the signals--- while the scheme in~\cite{lu2023experimental} is fully passive, containing no active elements at either the optical or the electronic level.
Secondly, the leakage system considered in the case of the modulator-free transmitter in~\cite{lo2023simplified} differs significantly from that of the passive transmitter. The number of optical modes per round that we use here to describe the leakage system is two, in contrast to three for the fully passive transmitter. One another note, the state of the leakage of the transmitter in~\cite{lo2023simplified} is identical for all signals with intensity $I_0$, as mentioned in~\cref{sec:OILtransmitter}. This may also explain the lower sensitivity to the information leakage.

\section{Conclusions}\label{sec:conclusions}
In this work, we have introduced a security analysis for decoy-state QKD schemes that use modulator-free transmitters in the presence of information leakage. To illustrate the method, we have applied it to two  prominent examples of such transmitters that inherently leak information about the encoded setting choices. We have used this analysis to evaluate the protocols' performance based on the attenuation provided by an intensity modulator that is placed inside the transmitter to mitigate the leakage. Our results show that the secret-key rate approaches that of an ideal leakage-free scenario when the attenuation provided by this device exceeds $\sim 70$ dB but drops sharply at low attenuation levels ($\lesssim 30$ dB). This highlights the need for strongly attenuating the undesired pulses in modulator-free transmitters as well as the importance of accounting for this type of side channel in QKD security proofs.

\section{Acknowledgments}
This work was supported by the Galician Regional Government (consolidation of Research Units: AtlantTIC), the Spanish Ministry of Economy and Competitiveness (MINECO), the Fondo Europeo de Desarrollo Regional (FEDER) through the grant No. PID2020-118178RB-C21, MICIN with funding from the European Union NextGenerationEU (PRTR-C17.I1) and the Galician Regional Government with own funding through the “Planes Complementarios de I+D+I con las Comunidades Autónomas” in Quantum Communication, the European Union’s Horizon Europe Framework Programme under the Marie Sklodowska-Curie Grant No. 101072637 (Project QSI) and the project ”Quantum Security Networks Partnership” (QSNP, grant agreement No 101114043).

\appendix

\section{Inverting target variables}\label{appendix:transformations}
The relations in~\cref{eq:transformations1} allow to precisely determine the target random variables $\theta$, $\phi$, and $\mu$, given the original random phases $\phi_1,\cdots,\phi_4$. However, the map is not bijective, and so one cannot determine the original phases given the target variables. To be precise, $\theta$ and $\mu$ determine the intensities
\begin{equation}
\mu_e=\mu\cos^2(\theta/2),  \qquad\qquad  \mu_l=\mu\sin^2(\theta/2),
\end{equation}
and, given $\mu_e$ and $\mu_l$, the value of the cosines
\begin{equation}\label{eq:cosines}
\cos(\phi_1-\phi_2)=\frac{2\mu_e}{\mu_{\rm max}}-1,  \qquad\qquad  \cos(\phi_3-\phi_4)=\frac{2\mu_l}{\mu_{\rm max}}-1,
\end{equation}
can be also determined. However, there are multiple values of the phase differences $\phi_1-\phi_2$ and $\phi_3-\phi_4$ that satisfy~\cref{eq:cosines}. 

Now, from the definition of $\phi_e$ in~\cref{eq:transformations1}, it is clear that the phase $e^{i\phi_e}$ is exactly halfway between $e^{i\phi_1}$ and $e^{i\phi_2}$, and the relative angle between the latter is determined by $\arccos(\frac{2\mu_e}{\mu_{\rm max}}-1)$. In particular, $e^{i\phi_e}$ is invariant under the exchange of $\phi_1$ and $\phi_2$, such that the occurrence of a certain value has two equally likely contributions. Analogously, the phases $\phi_3$ and $\phi_4$ are determined given $\mu_l$ and $\phi_l=\phi+\phi_e$. This leads to the four possible transformations that have been introduced in~\cref{eq:substitutions}.

\section{Quantum-coin argument}\label{appendix:quantum-coin-argument}
Here we include, for completeness, the derivation of the bounds used in \cref{eq:LPYields,eq:LPbiterror,eq:phase_error_bound1} based on the quantum-coin idea~\cite{gottesman,Lo-Preskill,Lo-Ma}. For simplicity we focus on the protocol with the fully passive transmitter introduced in~\cref{sec:PassiveTransmitter}, as the calculations for the other transmitter are entirely analogous.

\subsection{Yields}\label{appendix:quantum-coin-argument-Yields}
To relate the yields associated with different intensity settings, we consider that some of the rounds in which Alice prepares $\rho_{\abasis,I_0}^{\omega,n}$, $\rho_{\abasis,I_1}^{\omega,n}$ and $\rho_{\abasis,I_2}^{\omega,n}$ are substituted by the preparation of the entangled states
\begin{equation}\label{eq:coinIJ}
\begin{split}
    \ket*{\Psi_{IJ\text{-coin}}^{\omega,\beta,n}}_{CST}&=\frac{1}{\sqrt{2}}\left(\ket{0_{\beta}}_C\ket*{\rho_{\beta,I}^{\omega,n}}_{ST}+e^{i\varphi}\ket{1_{\beta}}_C\ket*{\rho_{\beta,J}^{\omega,n}}_{ST}\right),
\end{split}
\end{equation}
where $\ket*{\rho_{\abasis,I}^{\omega,n}}_{ST}$ is a purification of $\rho_{\abasis,I}^{\omega,n}$, with $I,J\in\bar{I}$ and $I\neq J$. Precisely, we can consider a fictitious scenario where each round in which Alice originally emits the state $\ket*{\rho_{\abasis,I}^{\omega,n}}_{ST}$ ($\ket*{\rho_{\abasis,J}^{\omega,n}}_{ST}$) is substituted by the preparation of the entangled state in~\cref{eq:coinIJ} with probability $p_{IJ\text{-coin}|I}^{\omega,\abasis,n}= (p_{IJ\text{-coin}}^{\omega,\abasis,n}/2)/(p_{n|\Omega_{\abasis}^{I}}^{\omega}p_{\Omega_{\abasis}^{I}})$ ($p_{IJ\text{-coin}|J}^{\omega,\abasis,n}= (p_{IJ\text{-coin}}^{\omega,\abasis,n}/2)/(p_{n|\Omega_{\abasis}^{J}}^{\omega}p_{\Omega_{\abasis}^{J}})$) for some prefixed probability $p_{IJ\text{-coin}}^{\omega,\abasis,n}<\min(p_{n|\Omega_{\abasis}^{I}}^{\omega}p_{\Omega_{\abasis}^{I}},p_{n|\Omega_{\abasis}^{J}}^{\omega}p_{\Omega_{\abasis}^{J}})$. Note that, if Alice measures the coin system $C$ in the $Z$ basis we recover the actual setting. In fact, we can consider a fictitious scenario for each combination $(I_0,I_1)$, $(I_0,I_2)$, and $(I_1,I_2)$. As we shall see below, the statistics of these $(IJ,\abasis,n)$-coin rounds in their corresponding fictitious scenarios must satisfy certain constraints that are used for parameter estimation. Importantly, as we focus here in the regime of infinitely many rounds, the specific values of the probabilities $p_{IJ\text{-coin}}^{\omega,\abasis,n}$ are not relevant and in fact they could be made as small as desired. Naturally, these assignment probabilities acquire more relevance in the finite-key regime, where one aims to improve the concentration bounds that relate different statistical sets (see \cite{Guille_framework}).

Now, from the Bloch sphere bound we have that the qubit system $C$ must satisfy the statistical relation $\langle \hat{Z}\rangle^2+\langle \hat{X}\rangle^2\leq 1$, where $\hat{Z}=\ketbra{0_Z}{0_Z}-\ketbra{1_Z}{1_Z}$ and $\hat{X}=\ketbra{0_X}{0_X}-\ketbra{1_X}{1_X}$. This means that $\left(1-2\Pr[1_Z^C|Z_C]\right)^2+\left(1-2\Pr[1_X^C|X_C]\right)^2\leq 1$, or equivalently~\cite{Lo-Preskill}
\begin{equation}\label{eq:Bloch}
  1-2\Pr[1_X^C|X_C]\leq2\sqrt{\Pr[1_Z^C|Z_C]\left(1-\Pr[1_Z^C|Z_C]\right)},
\end{equation}
where $a_{\beta}^C$ is the event "Alice obtains the outcome $a_{\beta}$ after measuring the coin system $C$", and ${\beta}_C$ is the event "Alice measures the coin system $C$ in the $\beta$ basis". Now suppose that before Alice measures the coin, Bob performs a measurement on the transmitted system $T$ with possible outcomes \textit{detected} (de) and \textit{undetected} (un). The former refers to the observation of a detection click in his measurement device, while the latter refers to a no-click event. This divides the rounds in two sets ---detected and undetected rounds--- and \cref{eq:Bloch} must be satisfied for each of these sets. That is, for each $b\in\set{\text{de},\text{un}}$ we have that
\begin{equation}\label{eq:Bloch2}
\begin{split}
  1-2\Pr[1_X^C|X_C,b]&\leq2\sqrt{\Pr[1_Z^C|Z_C,b]\left(1-\Pr[1_Z^C|Z_C,b]\right)}.
\end{split}
\end{equation}
Besides, since the probability of measuring the coin in the $X$ or $Z$ basis is fixed and independent of the outcome $b$, we have that
\begin{equation}
    \Pr[b]\Pr[1_X^C|X_C,b]=\Pr[b|X_C]\Pr[1_X^C|X_C,b]=\Pr[1_X^C,b|X_C],
\end{equation}
and similarly, 
\begin{equation}
\begin{split}
    \Pr[b]\Pr[1_Z^C|Z_C,b]&=\Pr[1_Z^C,b|Z_C],\\
    \Pr[b]\Pr[0_Z^C|Z_C,b]&=\Pr[0_Z^C,b|Z_C].
\end{split}
\end{equation}
Then, by multiplying both sides of~\cref{eq:Bloch2} by the probability $\Pr[b]$ one obtains
\begin{equation}
\begin{split}
    \Pr[b]-2\Pr[b]\Pr[1_X^C|X_C,b]&\leq2\sqrt{\Pr[1_Z^C|Z_C,b]\Pr[0_Z^C|Z_C,b]\Pr[b]^2}\\
    \Leftrightarrow  \Pr[b]-2\Pr[1_X^C,b|X_C]&\leq 2\sqrt{\Pr[1_Z^C,b|Z_C]\Pr[0_Z^C,b|Z_C]}.
\end{split}
\end{equation}
Finally, by adding the previous expressions for the two possible events $b\in\set{\text{de},\text{un}}$, one obtains
\begin{equation}\label{eq:quantum-coin-ineq1}
\begin{split}
    1-2\Pr[1_X^C|X_C] \leq & 2\sqrt{\Pr[\text{de}|1_Z^C,Z_C]\Pr[1_Z^C|Z_C]\Pr[\text{de}|0_Z^C,Z_C]\Pr[0_Z^C|Z_C]}\\
    &+2\sqrt{\Pr[\text{un}|1_Z^C,Z_C]\Pr[1_Z^C|Z_C]\Pr[\text{un}|0_Z^C,Z_C]\Pr[0_Z^C|Z_C]}\\
    =&2\sqrt{\Pr[0_Z^C|Z_C]\Pr[1_Z^C|Z_C]}\left(\sqrt{\Pr[\text{de}|1_Z^C,Z_C]\Pr[\text{de}|0_Z^C,Z_C]}+\sqrt{\Pr[\text{un}|1_Z^C,Z_C]\Pr[\text{un}|0_Z^C,Z_C]}\right)\\
    \leq& \sqrt{\Pr[\text{de}|1_Z^C,Z_C]\Pr[\text{de}|0_Z^C,Z_C]}+\sqrt{\Pr[\text{un}|1_Z^C,Z_C]\Pr[\text{un}|0_Z^C,Z_C]}\\
    =& \sqrt{Y_{\beta,I}^{\omega,n}Y_{\beta,J}^{\omega,n}}+\sqrt{(1-Y_{\beta,I}^{\omega,n})(1-Y_{\beta,J}^{\omega,n})}.
\end{split}
\end{equation}
Importantly, this inequality is satisfied for each of the quantum-coin rounds.

Now, if the purifications in~\cref{eq:coinIJ} are chosen such that $F_{I,J}^{\abasis,n}:=F(\rho_{\abasis,I}^{\omega,n},\rho_{\abasis,J}^{\omega,n})=\abs*{\braket*{\rho_{\abasis,I}^{\omega,n}}{\rho_{\abasis,J}^{\omega,n}}}^2$, then 
\begin{equation}\label{eq:coinIJ_derivation}
\begin{split}
    \sqrt{F_{I,J}^{\abasis,n}}
    =\abs*{\braket*{\rho_{\abasis,I}^{\omega,n}}{\rho_{\abasis,J}^{\omega,n}}}
    =\text{Re}\set{e^{i\varphi}\braket*{\rho_{\abasis,I}^{\omega,n}}{\rho_{\abasis,J}^{\omega,n}}}
    =1-2\Delta_{IJ},
\end{split}
\end{equation}
where $\Delta_{IJ}\equiv\Pr[1_X^C|X_C]=\norm*{ _C\braket*{1_X}{\Psi_{IJ\text{-coin}}^{\omega,\abasis,n}}_{CST}}_1$ is the quantum-coin imbalance~\cite{gottesman,Lo-Preskill} of the $(IJ,\abasis,n)$-coin. In~\cref{eq:coinIJ_derivation}, the second equality always holds for some value of $\varphi$, and the third equality follows from the definition in~\cref{eq:coinIJ}. Thus, combining~\cref{eq:coinIJ_derivation,eq:quantum-coin-ineq1}, we have
\begin{equation}\label{eq:coinIJ_derivation_final}
    \sqrt{F_{I,J}^{\abasis,n}}
    \leq \sqrt{Y_{\abasis,I}^{\omega,n}Y_{\abasis,J}^{\omega,n}}
    +\sqrt{(1-Y_{\abasis,I}^{\omega,n})(1-Y_{\abasis,J}^{\omega,n})},
\end{equation}
Note that the yields $Y_{\abasis,I}^{\omega,n}$ and $Y_{\abasis,J}^{\omega,n}$ associated with the $(IJ,\abasis,n)$-coin rounds are equal to the yields $Y_{\abasis,I}^{\omega,n}$ and $Y_{\abasis,J}^{\omega,n}$ of the remaining set of $(I,\abasis,n)$- and $(J,\abasis,n)$-rounds. \cref{eq:coinIJ_derivation_final} can be solved for one of the yields, obtaining
\begin{equation}\label{eq:GfunctionAppendix}
    G_{F_{I,J}^{\abasis,n}}^{\rm L}\left(Y_{\abasis,J}^{\omega,n}\right)\leq Y_{\abasis,I}^{\omega,n} \leq G_{F_{I,J}^{\abasis,n}}^{\rm U}\left(Y_{\abasis,J}^{\omega,n}\right),
\end{equation}
where $G_{z}^{K}(y)$ is defined in~\cref{sec:security}.

Finally, one can always take linear approximations of the functions $G_{z}^{K}(y)$ at some reference points $\tilde{y}$ to use these constraints in a linear program~\cite{zapatero2021security}, as shown in~\cref{eq:LPYields}. Since $G_{z}^{\rm L}(y)$ ($G_{z}^{\rm U}(y)$) is a convex (concave) function, these linear approximations lead to relaxations of the original non-linear optimization problem, and so the bound computed by the linear program is valid.

%%%%%%%%%%%%%%%
\subsection{Phase-error rate}\label{appendix:quantum-coin-imbalance-ph}
We start from
\begin{equation}\label{eq:CoinZXstate}
\begin{split}
    \ket{\Psi_{ZX\text{-coin}}^{\omega}}_{CAST}&=\frac{1}{\sqrt{2}}\left(\ket{0}_C \ket*{\psi_{Z,I_0}^{\omega,1}}_{AST}+\ket{1}_C\ket*{\psi_{X,I_0}^{\omega,1}}_{AST}\right),
\end{split}
\end{equation}
which represents the state that Alice prepares in the so-called single-photon $ZX$-coin rounds, where $\ket*{\psi_{\beta,I_0}^{\omega,1}}$ is given in~\cref{eq:virtual_state}, \textit{i.e.}, is a purification of the state $\rho_{\beta,I_0}^{\omega,1}$ transmitted in the single-photon $\Omega_{\beta}^{I_0}$ rounds. Similar to the previous section, one can assume that, with probability $p_{ZX\text{-coin}|\abasis}^{I_0,1}=(p_{ZX\text{-coin}}^{I_0,1}/2)/(p_{1|\Omega_{\abasis}^{I_0}}p_{\abasis,I_0})$, for some $p_{ZX\text{-coin}}^{I_0,1}<\min(p_{1|\Omega_{Z}^{I_0}}p_{Z,I_0},p_{1|\Omega_{X}^{I_0}}p_{X,I_0})$, the $\ket*{\psi_{\beta,I_0}^{\omega,1}}$ round is substituted by the preparation of the coin state given by~\cref{eq:CoinZXstate}. Again, these rounds satisfy certain constraints that are used in the parameter estimation. 

Importantly, as in~\cref{appendix:quantum-coin-argument-Yields}, we have that the coin system $C$ must satisfy~\cref{eq:Bloch2} for any measurement outcome $b$. In particular, in this context we consider that, before Alice measures the coin system $C$, Alice and Bob measure systems $AT$ in the $X$ basis, obtaining either the outcome \textit{same} (sa) ---when their outcomes are equal--- \textit{error} (er) ---when they are not equal--- or \textit{undetected} (un). This divides the $ZX$-coin rounds in three sets. Still, it is clear that the Bloch-sphere bound provided by~\cref{eq:Bloch2} must be satisfied for each of them. Besides, the probability of measuring the coin system $C$ in the $X_C$ or $Z_C$ basis is again fixed and independent of the outcome $b\in\set{\text{sa},\text{er},\text{un}}$. This means that, like in~\cref{appendix:quantum-coin-argument-Yields}, we can take the sum of the Bloch-sphere bound for the cases "sa" and "er", and, since $\text{de}\Leftrightarrow\text{sa}\vee\text{er}$, where "de" refers to the \textit{detected} outcome, we obtain
\begin{equation}
\begin{split}
    \Pr[\text{de}]-2\Pr[1_X^C,\text{de}|X_C] \leq 
    &2\sqrt{\Pr[0_Z^C|Z_C]\Pr[1_Z^C|Z_C]}\\
    &\times\left(\sqrt{\Pr[\text{ph}|1_Z^C,Z_C]\Pr[\text{ph}|0_Z^C,Z_C]}+\sqrt{\Pr[\text{sa}|1_Z^C,Z_C]\Pr[\text{sa}|0_Z^C,Z_C]}\right)\\
    \leq& \sqrt{\Pr[\text{er}|1_Z^C,Z_C]\Pr[\text{er}|0_Z^C,Z_C]}+\sqrt{\Pr[\text{sa}|1_Z^C,Z_C]\Pr[\text{sa}|0_Z^C,Z_C]}.
\end{split}
\end{equation}
Finally, we can divide both sides by the detection probability in a $ZX$-quantum-coin round, that we shall denote by $\Pr[{\rm de}]\equiv Y_{ZX\text{-coin}}$, and use the inequality $\Pr[1_X^C,\text{de}|X_C]\leq \Pr[1_X^C|X_C]$, to obtain
\begin{equation}\label{eq:coin_ineq_ph}
\begin{split}
    1-2\frac{\Pr[1_X^C|X_C]}{Y_{ZX\text{-coin}}} \leq 
    & \sqrt{e_{\rm ph}^{\omega}e_{X,I_0}^{\omega,1}}+\sqrt{(1-e_{\rm ph}^{\omega})(1-e_{X,I_0}^{\omega,1})}.
\end{split}
\end{equation}
This inequality is satisfied for each of the $ZX$-quantum-coin rounds. 

The probability $\Pr[1_X^C|X_C]\equiv\Delta_{ZX}$ in a $ZX$-quantum-coin round ---which is required to make use of~\cref{eq:coin_ineq_ph} in our security analysis--- can be computed as
\begin{equation}\label{eq:QCimbalanceZX}
    \Delta_{ZX}:=\frac{1-\text{Re}\set{\braket*{\psi_{Z,I_0}^{\omega,1}}{\psi_{X,I_0}^{\omega,1}}}}{2},
\end{equation}
where the term $\text{Re}\set{\braket*{\psi_{Z,I_0}^{\omega,1}}{\psi_{X,I_0}^{\omega,1}}}$ is calculated in~\cref{appendix:quantum-coin-imbalance}. Finally, we define the quantity $F_{ZX}':=(1-2\Delta_{ZX}/Y_{ZX\text{-coin}}^{\rm L})^2$ and solve~\cref{eq:coin_ineq_ph}, similarly to~\cref{eq:coinIJ_derivation,eq:GfunctionAppendix}, to obtain an upper bound on the phase error rate
\begin{equation}
    e_{\rm ph}^{\omega}\leq G_{F_{ZX}'}^{\rm U}\left(e_{X,I_0}^{\omega,1}\right),
\end{equation}
Again, we still obtain a valid bound if we substitute $Y_{ZX\text{-coin}}$ ($e_{X,I_0}^{\omega,1}$) by its corresponding lower (upper) bound $Y_{ZX\text{-coin}}^{\rm L}=(Y_{Z,I_0}^{\omega,1,{\rm L}}+Y_{X,I_0}^{\omega,1,{\rm L}})/2$ ($e_{X,I_0}^{\omega,1,{\rm U}}$).

\section{Efficient fidelity estimation}\label{appendix:EfficientFidelity}
To solve the linear programs presented in the main text, we need to estimate the fidelities $F_{I,J}^{\beta,n}\equiv F(\rho_{\beta,I}^{\omega,n},\rho_{\beta,J}^{\omega,n})$ defined in~\cref{subsec:single_photon_yield}. The matrices $\rho_{\beta,I}^{\omega,n}$ can be numerically computed, which means that, in principle, one could also compute the desired fidelities directly. Unfortunately, even for moderate values of $n$, the size of these matrices makes this process very slow. Note that, for each $n$ and for each pair of intensities, first one needs to compute the matrices ---for which one has to solve a triple integral for the ${n+4\choose n}^2$ entries of each of the two matrices involved--- and then calculate the fidelity.

To speed up this process, one can estimate these fidelities from a sub-block of each matrix. For this, let us consider that we sort the rows and columns of each $n$-photon matrix such that those associated with vectors $\ket{n_e,n_l,n_1,n_3,n_5}_{el135}$ with lower $n_L\equiv n_1+n_3+n_5$ go first. Then, we could neglect all the entries of the matrix for which $n_L>n_L^{\rm cut}$ for some $n_L^{\rm cut}$. For instance, if we simply set $n_L^{\rm cut}=0$, the first sub-block of the matrix corresponds to the entries for which all the leakage systems are in the vacuum state (\textit{i.e.}, entries associated with the states $\ket{n_e,n_l,0,0,0}_{135}$ with $n_e+n_l=n$). An important insight here is that the remaining entries of the matrix that are outside of this sub-block must be close to zero (as the intensity $\mu_L$ of the leakage system $L$ is typically very small). Similarly, if we set $n_L^{\rm cut}=1$, this implies that we consider a sub-block containing the entries associated with the states $\ket{n_e,n_l,n_1,n_3,n_5}_{el135}$ that satisfy $n_e+n_l+n_1+n_3+n_5=n$ and $n_1+n_3+n_5\leq 1$, being all the remaining entries neglected. In doing so, we can obtain a lower bound on the desired fidelity. For this, we use the following results~\cite{curras2023security,sixto2023secret}:
\begin{result}[Fidelity with a projection~\cite{curras2023security}]\label{res:fid_proj}
     Let $\rho$ be a density matrix, and let $\rho'=\frac{\Pi\rho\Pi}{\textnormal{Tr}[\Pi\rho\Pi]}$, where $\Pi$ is a projector. Then $F(\rho,\rho')=\textnormal{Tr}[\Pi\rho\Pi]$.
\end{result}
\begin{result}[Bures-distance-based fidelity bound]\label{res:bures}
     Let $\rho$, $\sigma$, and $\tau_i$ ($i=1,\dots,N$) be density operators. Then, the fidelity $F(\rho,\sigma)$ satisfies
     \begin{equation}
         \sqrt{F(\rho,\sigma)}\geq 1-\frac{1}{2}\left[d_B(\rho,\tau_1)+d_B(\tau_1,\tau_2)+\dots+d_B(\tau_{N-1},\tau_N)+d_B(\tau_N,\sigma)\right]^2.
     \end{equation}
     where $d_B(\rho,\sigma):=\sqrt{2\left(1-\sqrt{F(\rho,\sigma)}\right)}$ is the Bures distance between $\rho$ and $\sigma$.
\end{result}
\cref{res:bures} straightforwardly follows from the definition of the Bures distance and the triangle inequality.

The idea is to reduce the problem of calculating the fidelity $F_{I,J}^{\beta,n}$ to the computationally simpler problem of computing $F_{I,J,\Pi}^{\beta,n}\equiv F(\rho_{\beta,I,\Pi}^{\omega,n},\rho_{\beta,J,\Pi}^{\omega,n})$, where the operator $\Pi$ that appears in the subscripts indicates that the density operator is a normalized projection onto the desired subspace in which the leakage system contains $n_L^{\rm cut}$ photons or less. For instance, if $n_L^{\rm cut}=0$, we have that $\Pi=\identity_{el}^{(n)}\otimes\dyad*{\rm vac}_{135}$, where $\identity_{el}^{(n)}=\sum_{m=0}^n\dyad{m,n-m}_{el}$ is the identity matrix in the $n$-photon subspace associated with the optical modes $el$. Then, we calculate the projected operators as
\begin{equation}
\begin{split}
    \rho_{\beta,I,\Pi}^{\omega,n}&=\frac{\Pi\rho_{\beta,I}^{\omega,n}\Pi}{\textnormal{Tr}[\Pi\rho_{\beta,I}^{\omega,n}\Pi]}
    =\frac{\Pi\langle \bar\sigma_{\theta,\phi,\mu}^{\omega,n}\rangle_{\Omega_{\abasis}^{I}}\Pi}{\textnormal{Tr}[\Pi\langle\bar\sigma_{\theta,\phi,\mu}^{\omega,n}\rangle_{\Omega_{\abasis}^{I}}\Pi]}
    =\frac{\langle\Pi \bar\sigma_{\theta,\phi,\mu}^{\omega,n}\Pi\rangle_{\Omega_{\abasis}^{I}}}{\textnormal{Tr}[\langle\Pi\bar\sigma_{\theta,\phi,\mu}^{\omega,n}\Pi\rangle_{\Omega_{\abasis}^{I}}]}
    =\frac{\langle\bar\sigma_{\theta,\phi,\mu,\Pi}^{\omega,n}\rangle_{\Omega_{\abasis}^{I}}}{\textnormal{Tr}[\langle\bar\sigma_{\theta,\phi,\mu,\Pi}^{\omega,n}\rangle_{\Omega_{\abasis}^{I}}]},
\end{split}
\end{equation}
and the trace
\begin{equation}\label{eq:traceLemmaUtility}
    \textnormal{Tr}[\Pi\rho_{\beta,I}^{\omega,n}\Pi]=\frac{\textnormal{Tr}[\langle\bar\sigma_{\theta,\phi,\mu,\Pi}^{\omega,n}\rangle_{\Omega_{\abasis}^{I}}]}{p_{n|\Omega_{\abasis}^{I},\omega}\langle 1\rangle_{\Omega_{\abasis}^{I}}}.
\end{equation}
Note that, in this example, the operators $\langle\bar\sigma_{\theta,\phi,\mu,\Pi}^{\omega,n}\rangle_{\Omega}$ live in a $(n+1)$-dimensional space, and so they are easy to compute for relatively small $n$. Also note that 
\begin{equation}
    p_{n|\Omega,\omega}\langle1\rangle_{\Omega}=\Tr{\langle \bar\sigma_{\theta,\phi,\mu}^{\omega,n}\rangle_{\Omega}}=\Big\langle\Tr{ \bar\sigma_{\theta,\phi,\mu}^{\omega,n}}\Big\rangle_{\Omega}=\Big\langle e^{-\mu-\mu_L}\frac{(\mu+\mu_L)^n}{n!}\Big\rangle_{\Omega}.
\end{equation}
Finally, from~\cref{res:bures}, we have that
\begin{equation}
    F(\rho_{\beta,I}^{\omega,n},\rho_{\beta,J}^{\omega,n})
    \geq \left[1-\frac{1}{2}\left(d_B(\rho_{\beta,I}^{\omega,n},\rho_{\beta,I,\Pi}^{\omega,n})
    +d_B(\rho_{\beta,I,\Pi}^{\omega,n},\rho_{\beta,J,\Pi}^{\omega,n})
    +d_B(\rho_{\beta,J,\Pi}^{\omega,n},\rho_{\beta,J}^{\omega,n})\right)^2\right]^2,
\end{equation}
where $d_B(\rho_{\beta,I}^{\omega,n},\rho_{\beta,I,\Pi}^{\omega,n})$ and $d_B(\rho_{\beta,J,\Pi}^{\omega,n},\rho_{\beta,J}^{\omega,n})$ can be obtained from~\cref{eq:traceLemmaUtility} via~\cref{res:fid_proj}.
\section{Inner product between purifications}\label{appendix:quantum-coin-imbalance}
To compute $\text{Re}\set{\braket*{\psi_{Z,I_0}^{\omega,1}}{\psi_{X,I_0}^{\omega,1}}}$ we consider specific purifications for both $\ket*{\psi_{Z,I_0}^{\omega,1}}_{AST}$ and $\ket*{\psi_{X,I_0}^{\omega,1}}_{AST}$ (see~\cref{eq:virtual_state}), which depend on the purifications
\begin{equation}\label{eq:purificationBasis}
    \ket*{\rho_{\abit,\abasis,I_0}^{\omega,1}}_{ST}=\sum_{j=0}^{d-1}\sqrt{q_{\abit,\abasis,j}}e^{i\xi_{\abit,\abasis,j}}\ket{j}_S\otimes\ket*{\varphi_{\abit,\abasis,j}}_T,
\end{equation}
where the state $\ket*{\varphi_{\abit,\abasis,j}}$ is the eigenvector of $\rho_{\abit,\abasis,I_0}^{\omega,1}$ associated with its $(j+1)$-th largest eigenvalue $q_{\abit,\abasis,j}$, and $\xi_{\abit,\abasis,j}\in(-\pi,\pi]$ are arbitrary phases. For instance, these quantities can be obtained numerically from $\rho_{\abit,\abasis,I_0}^{\omega,1}$. We note that in the ideal scenario without information leakage (\textit{i.e.}, $\omega=0$), we can simply select the following purification
\begin{equation}
\begin{split}
    \ket*{\rho_{0,Z,I_0}^{\omega=0,1}}_{ST}&=\sqrt{q_{0,Z}}\ket{0}_S\ket*{0_Z}_T +e^{i\xi_{0,Z}}\sqrt{1-q_{0,Z}}\ket{1}_S\ket*{1_Z}_T,\\
    \ket*{\rho_{1,Z,I_0}^{\omega=0,1}}_{ST}&=\sqrt{q_{1,Z}}\ket{0}_S\ket*{1_Z}_T +e^{i\xi_{1,Z}}\sqrt{1-q_{1,Z}}\ket{1}_S\ket*{0_Z}_T,\\
    \ket*{\rho_{0,X,I_0}^{\omega=0,1}}_{ST}&=\sqrt{q_{0,X}}\ket{0}_S\ket*{0_X}_T +e^{i\xi_{0,X}}\sqrt{1-q_{0,X}}\ket{1}_S\ket*{1_X}_T,\\
    \ket*{\rho_{1,X,I_0}^{\omega=0,1}}_{ST}&=\sqrt{q_{1,X}}\ket{0}_S\ket*{1_X}_T +e^{i\xi_{1,X}}\sqrt{1-q_{1,X}}\ket{1}_S\ket*{0_X}_T,
\end{split}
\end{equation}
where we have set $\xi_{\abit,\abasis,0}=0$ for all $\abit,\abasis$, and we have omitted the index $j=1$ in the phases $\xi_{\abit,\abasis,1}$ for simplicity of notation. Thus,
\begin{equation}
\begin{split}
    \braket*{\psi_{Z,I_0}^{\omega=0,1}}{\psi_{X,I_0}^{\omega=0,1}}
    &=\frac{1}{2\sqrt{2}}\left[\braket*{\rho_{0,Z,I_0}^{\omega=0,1}}{\rho_{0,X,I_0}^{\omega=0,1}} + \braket*{\rho_{0,Z,I_0}^{\omega=0,1}}{\rho_{1,X,I_0}^{\omega=0,1}}
    +\braket*{\rho_{1,Z,I_0}^{\omega=0,1}}{\rho_{0,X,I_0}^{\omega=0,1}} - \braket*{\rho_{1,Z,I_0}^{\omega=0,1}}{\rho_{1,X,I_0}^{\omega=0,1}}\right]\\
    &=\frac{1}{4}\left[\sqrt{q_{0,Z}q_{0,X}}+\sqrt{q_{0,Z}q_{1,X}}+\sqrt{q_{1,Z}q_{0,X}}+\sqrt{q_{1,Z}q_{1,X}}\right]\\
    &+\frac{1}{4}\left[-e^{i(\xi_{0,X}-\xi_{0,Z})}\sqrt{(1-q_{0,Z})(1-q_{0,X})} +e^{i(\xi_{1,X}-\xi_{0,Z})}\sqrt{(1-q_{0,Z})(1-q_{1,X})}\right.\\
    &+\left. e^{i(\xi_{0,X}-\xi_{1,Z})}\sqrt{(1-q_{1,Z})(1-q_{0,X})} -e^{i(\xi_{1,X}-\xi_{1,Z})}\sqrt{(1-q_{1,Z})(1-q_{1,X})}\right]\\
    &=\frac{1}{4}\left[\sqrt{q_{0,Z}q_{0,X}}+\sqrt{q_{0,Z}q_{1,X}}+\sqrt{q_{1,Z}q_{0,X}}+\sqrt{q_{1,Z}q_{1,X}}\right]\\
    &+\frac{1}{4}\left[\sqrt{(1-q_{0,Z})(1-q_{0,X})} +\sqrt{(1-q_{0,Z})(1-q_{1,X})}\right.\\
    &+\left.\sqrt{(1-q_{1,Z})(1-q_{0,X})} +\sqrt{(1-q_{1,Z})(1-q_{1,X})}\right],
\end{split}
\end{equation}
where we have set $\xi_{0,Z}=\xi_{1,X}=\pi$ and $\xi_{1,Z}=\xi_{0,X}=0$ (we could have equivalently chosen $\xi_{0,Z}=\xi_{1,X}=0$ and $\xi_{1,Z}=\xi_{0,X}=\pi$). Note that when $q_{\abit,\abasis}=q$, we obtain $ \braket*{\psi_{Z,I_0}^{\omega=0,1}}{\psi_{X,I_0}^{\omega=0,1}}=1$.

In the more general setting where $\omega>0$, one expects the intensity of the leakage systems to be small, and so one can use a similar purification like the one above. That is, one could fix $\xi_{0,Z,1}=\xi_{1,X,1}=\pi$ and set $\xi_{a,\beta,j}=0$ for any other values of $\abit$, $\abasis$ and $j$. In addition, note that we expect the quantities $q_{\abit,\abasis,j}$ in~\cref{eq:purificationBasis} to be very small for $j\geq2$, so the complex phases associated with these terms are not really important. Still, one could always optimize all of these parameters to tighten the bounds. Thus, we can compute the inner product as
\begin{equation}
\begin{split}
    \braket*{\psi_{Z,I_0}^{\omega,1}}{\psi_{X,I_0}^{\omega,1}}
    =&\frac{1}{2\sqrt{2}}\left[\braket*{\rho_{0,Z,I_0}^{\omega,1}}{\rho_{0,X,I_0}^{\omega,1}} + \braket*{\rho_{0,Z,I_0}^{\omega,1}}{\rho_{1,X,I_0}^{\omega,1}}
    +\braket*{\rho_{1,Z,I_0}^{\omega,1}}{\rho_{0,X,I_0}^{\omega,1}} - \braket*{\rho_{1,Z,I_0}^{\omega,1}}{\rho_{1,X,I_0}^{\omega,1}}\right]\\
    =&\sum_{j=0}^{d-1}\left(\sqrt{q_{0,Z,j}q_{0,X,j}}e^{i(\xi_{0,X,j}-\xi_{0,Z,j})}\braket*{\varphi_{0,Z,j}}{\varphi_{0,X,j}}
    +\sqrt{q_{0,Z,j}q_{1,X,j}}e^{i(\xi_{1,X,j}-\xi_{0,Z,j})}\braket*{\varphi_{0,Z,j}}{\varphi_{1,X,j}}\right.\\
    &+\left.\sqrt{q_{1,Z,j}q_{0,X,j}}e^{i(\xi_{0,X,j}-\xi_{1,Z,j})}\braket*{\varphi_{1,Z,j}}{\varphi_{0,X,j}}
    -\sqrt{q_{1,Z,j}q_{1,X,j}}e^{i(\xi_{1,X,j}-\xi_{1,Z,j})}\braket*{\varphi_{1,Z,j}}{\varphi_{1,X,j}}\right),
\end{split}
\end{equation}
from where can we directly obtain $\text{Re}\set{\braket*{\psi_{Z,I_0}^{\omega,1}}{\psi_{X,I_0}^{\omega,1}}}$.
\section{Refined security analysis for a passive based on post-selection transmitter}\label{appendix:alternativeSec}
Even in the idealized scenario where the undesired pulses are completely blocked by the IM ---\textit{i.e.}, when $\omega = 0$--- the quantum states emitted by the passive transmitter remain intrinsically noisy due to the mixed nature of the post-selected states. This noise elevates the single-photon bit-error rate, ultimately undermining the protocol's performance.

To further enhance resulting secret-key rate, the security analysis introduced in~\cref{sec:security} can be improved for this particular type of noisy transmitters by considering a modified virtual scenario. Specifically, now, rather than generating the secret key from those detected rounds in which Alice emits $\rho_{0,Z,I_0}^{\omega,1}$ or $\rho_{1,Z,I_0}^{\omega,1}$ and Bob measures in the $Z$ basis, we shall consider that both users distill key from the subset of those rounds in which Alice emits the eigenvectors of $\rho_{0,Z,I_0}^{\omega,1}$ and $\rho_{1,Z,I_0}^{\omega,1}$ with the largest eigenvalues. Precisely, let us denote these states as $\ket*{\varphi_{0,Z,I_0}^{\omega,\rm{key}}}$ and $\ket*{\varphi_{1,Z,I_0}^{\omega,\rm{key}}}$, with $\ket*{\varphi_{a,\beta,I}^{\omega,\rm{key}}}$ being the eigenvector of $\rho_{a,\beta,I}^{\omega,1}$ associated with the largest eigenvalue $q_{a,\beta,I}^{\omega,\rm{key}}$. Besides, let us denote as $\ket*{\varphi_{a,\beta,I}^{\omega,\rm{opp}}}$ the eigenvector associated with the second largest eigenvalue $q_{a,\beta,I}^{\omega,\rm{opp}}$. We use the tag "opp" because this second eigenvector $\ket*{\varphi_{a,\beta,I}^{\omega,\rm{opp}}}$ points in the opposite direction of $\ket*{\varphi_{a,\beta,I}^{\omega,\rm{key}}}$ on the Bloch sphere when $\omega=0$. 

In this alternative scenario, the entangled virtual state of Alice's "key" emissions ---$\ket*{\varphi_{0,Z,I_0}^{\omega,\rm{key}}}$ or $\ket*{\varphi_{1,Z,I_0}^{\omega,\rm{key}}}$--- is given by
\begin{equation}\label{eq:virtual_state_key}
    \ket*{\psi_{Z,I_0}^{\omega,\rm{key}}}_{\SAnc\STra}=\frac{1}{\sqrt{2}}\left[\ket{0_{Z}}_{\SAnc}\ket*{\varphi_{0,Z,I_0}^{\omega,\rm{key}}}_{\STra}+\ket{1_{Z}}_{\SAnc}\ket*{\varphi_{0,Z,I_0}^{\omega,\rm{key}}}_{\STra}\right],
\end{equation} 
and therefore the secret-key rate formula must be adjusted as
\begin{equation}
    R\geq p_{Z_B}p_{\Omega^{I_0}_{Z}}p_{1|\Omega^{I_0}_{Z}}^{\omega}q_{Z,I_0}^{\omega,\rm{key}}Y_{Z,I_0}^{\omega,1,\rm{key},{\rm L}}\left[1-h_2\left(e_{\rm ph}^{\omega,1,{\rm key,U}}\right)\right]- p_{Z_B}p_{\Omega^{I_0}_{Z}} Q_{Z,I_0} f_{\rm EC} h_2(E_{Z,I_0}),
\end{equation}
where $q_{\beta,I}^{\omega,t}:=\frac{1}{2}[q_{0,\beta,I}^{\omega,t}+q_{1,\beta,I}^{\omega,t}]$, with $t\in\set{\rm{key},\rm{opp}}$, and the quantities $Y_{\beta,I}^{\omega,1,\rm{key}}$ and $e_{\rm ph}^{\omega,1,{\rm key}}$ are defined similarly to their equivalents introduced the main text, but now referring to the eigenvectors $\ket*{\varphi_{a,\beta,I}^{\omega,\rm{key}}}$ instead of to the single-photon mixed operator $\rho_{\beta,I}^{\omega,1}$ or the purifications $\ket*{\rho_{a,\beta,I}^{\omega,1}}$. A precise definition of these quantities is provided below.

Now, to obtain $Y_{Z,I_0}^{\omega,1,\rm{key},{\rm L}}$ (and also $Y_{X,I_0}^{\omega,1,\rm{key},{\rm L}}$), we note that the emissions of $\rho_{\abasis,I}^{\omega,1}=\frac{1}{2}(\rho_{0,\abasis,I}^{\omega,1}+\rho_{1,\abasis,I}^{\omega,1})$ in the actual protocol can be replaced by the emission of the states $\tau_{\abasis,I}^{\omega,t}:=\frac{1}{2}\left(\ketbra*{\varphi_{0,\beta,I}^{\omega,t}}+\ketbra*{\varphi_{1,\beta,I}^{\omega,t}}\right)$ with probability $q_{\beta,I}^{\omega,t}$, with $t\in\set{\rm{key},\rm{opp}}$, plus some other state $\tau_{\beta,I}^{\omega,\rm{rest}}\propto\rho_{\abasis,I}^{\omega,1}-q_{\beta,I}^{\omega,\rm{key}}\tau_{\abasis,I}^{\omega,\rm{key}}-q_{\beta,I}^{\omega,\rm{opp}}\tau_{\beta,I}^{\omega,\rm{opp}}$ emitted with probability $1-q_{\beta,I}^{\omega,\rm{key}}-q_{\beta,I}^{\omega,\rm{opp}}$. Also, note that for small $\omega$ we expect that $\tau_{\abasis,I}^{\omega,\rm{key}}\approx\tau_{\abasis,I}^{\omega,\rm{opp}}\approx\frac{1}{2} \identity_{el}$, with $\identity_{el}$ being a projector onto the subspace of single-photon states in the joint mode $el$. We can incorporate this information into the problem and construct the following LP:
\begin{gather}\label{eq:LP1appendix}
\begin{aligned} 
    \textup{min}\quad& Y_{\abasis,I_0}^{\omega,1,\rm{key}} \\
    \textup{s.t.}\quad 
    & \sum_{n=0}^{n_{\rm cut}}p_{n|\Omega_{\abasis}^{I}}^{\omega}Y_{\abasis,I}^{\omega,n} 
    \leq Q_{\abasis,I}\leq 
    1- \sum_{n=0}^{n_{\rm cut}}p_{n|\Omega_{\abasis}^{I}}^{\omega}(1-Y_{\abasis,I}^{\omega,n})\qquad (I \in \bar{I}),\\
    & \text{LCS}_{F_{I,J}^{\abasis,n}}^{\rm L}\left(Y_{\abasis,I}^{\omega,n}\right) \leq Y_{\abasis,J}^{\omega,n} \leq
    \text{LCS}_{F_{I,J}^{\abasis,n}}^{\rm U}\left(Y_{\abasis,I}^{\omega,n}\right),
    \qquad (n\leq n_{\rm cut}, I\neq J\in\bar{I}),\\
    & Y_{\abasis,I}^{\omega,1} \geq q_{\abasis,I}^{\omega,\rm{key}}Y_{\abasis,I}^{\omega,1,\rm{key}}+ q_{\abasis,I}^{\omega,\rm{opo}}Y_{\abasis,I}^{\omega,1,\rm{opp}},\qquad (I\in\bar{I}),\\
    & Y_{\abasis,I}^{\omega,1} \leq q_{\abasis,I}^{\omega,\rm{key}}Y_{\abasis,I}^{\omega,1,\rm{key}}+ q_{\abasis,I}^{\omega,\rm{opo}}Y_{\abasis,I}^{\omega,1,\rm{opp}} + (1-q_{\abasis,I}^{\omega,\rm{key}}-q_{\abasis,I}^{\omega,\rm{opo}}) ,\qquad (I\in\bar{I}),\\
    & \text{LCS}_{F_{I,J}^{\abasis,t}}^{\rm L}\left(Y_{\abasis,I}^{\omega,1,t}\right) 
    \leq Y_{\abasis,J}^{\omega,1,t} \leq
    \text{LCS}_{F_{I,J}^{\abasis,t}}^{\rm U}\left(Y_{\abasis,I}^{\omega,1,t}\right),
    \qquad  (I\neq J \in\bar{I},\quad t\in\set{{\rm key,opp}}),\\
    & \text{LCS}_{F_{t,t'}^{\abasis,I}}^{\rm L}\left(Y_{\abasis,I}^{\omega,1,t}\right) 
    \leq Y_{\abasis,I}^{\omega,1,t'} \leq 
    \text{LCS}_{F_{t,t'}^{\abasis,I}}^{\rm U}\left(Y_{\abasis,I}^{\omega,1,t}\right),
    \qquad  (I \in\bar{I},\quad t\neq t'\in\set{{\rm key,opp}}),\\
    & 0\leq Y_{\abasis,I}^{\omega,n} \leq 1
    \qquad (n\leq n_{\rm cut}, I\in\bar{I}),\\
    & 0\leq Y_{\abasis,I}^{\omega,1,t} \leq 1
    \qquad (t\in\{\rm{key,opp}\}),
\end{aligned}
\end{gather}
where $Y_{\abasis,I}^{\omega,1,t}$, with $t\in\set{{\rm key,opp}}$, is the yield associated with the state $\tau_{\abasis,I}^{\omega,t}$, and the fidelities $F_{I,J}^{\abasis,t}:=F(\tau_{\abasis,I}^{\omega,t},\tau_{\abasis,J}^{\omega,t})$ and $F_{t,t'}^{\abasis,I}:=F(\tau_{\abasis,I}^{\omega,t},\tau_{\abasis,I}^{\omega,t'})$ can be computed precisely as the states are fully characterized. Note that some constraints in the LP in~\cref{eq:LP1appendix} are identical to those included in the LP introduced in~\cref{eq:LPYields}. Importantly, now the objective function is the "key" yield $Y_{\abasis,I}^{\omega,\rm{key}}$. Moreover, the LP in~\cref{eq:LP1appendix} incorporates similarity constraints between the "key" yields associated with different intensity settings, as well as between "key" and "opp" yields.

Similarly, to estimate the $X$-basis bit-error rate, we can employ the following LP: 
\begin{gather}\label{eq:LP2appendix}
\begin{aligned} 
    \textup{max}\quad& \Gamma_{X,I_0}^{\omega,1,\rm{key}}:= \frac{1}{2}\left(Y_{0,X,I_0}^{1,\omega,1,\rm{key}}+Y_{1,X,I_0}^{0,\omega,1,\rm{key}}\right) \\
    \textup{s.t.}\quad
    &\sum_{n=0}^{n_{\rm cut}}p_{n|\Omega_{\abit,X}^{I}}^{\omega}Y_{a,X,I}^{b,\omega,n} 
    \leq Q_{a,X,I}^{b}\leq 
    1- \sum_{n=0}^{n_{\rm cut}}p_{n|\Omega_{\abit,X}^{I}}^{\omega}(1-Y_{a,X,I}^{b,\omega,n}),\quad(a,b\in\set{0,1}, I\in\bar{I}),\\
    &\text{LCS}_{F_{I,J}^{a,X,n}}^{\rm U}\left(Y_{a,X,I}^{b,\omega,n}\right) 
    \leq Y_{a,X,J}^{b,\omega,n} \leq
    \text{LCS}_{F_{I,J}^{a,X,n}}^{\rm U}\left(Y_{a,X,I}^{b,\omega,n}\right),
    \quad (a,b\in\set{0,1}, n\leq n_{\rm cut},  I\neq J\in\bar{I}),\\
    &Y_{a,X,I}^{b,\omega,1} \geq q_{\abit,X,I}^{\omega,\rm{key}}Y_{a,X,I}^{b,\omega,1,\rm{key}}+ q_{\abit,X,I}^{\omega,\rm{opp}}Y_{a,X,I}^{b,\omega,1,\rm{opp}},\\
    &Y_{a,X,I}^{b,\omega,1} \leq q_{\abit,X,I}^{\omega,\rm{key}}Y_{a,X,I}^{b,\omega,1,\rm{key}}+ q_{\abit,X,I}^{\omega,\rm{opp}}Y_{a,X,I}^{b,\omega,1,\rm{opp}} +(1-q_{\abit,X,I}^{\omega,\rm{key}}-q_{\abit,X,I}^{\omega,\rm{opp}}),\qquad (a,b\in\set{0,1}, I\in\bar{I})\\
    & \text{LCS}_{F_{I,J}^{a,X,t}}^{\rm L}\left(Y_{a,X,I}^{b,\omega,1,t}\right) 
    \leq Y_{a,X,J}^{b,\omega,1,t} \leq
    \text{LCS}_{F_{I,J}^{a,X,t}}^{\rm U}\left(Y_{a,X,I}^{b,\omega,1,t}\right),
    \qquad (a,b\in\set{0,1}, I\neq J\in\bar{I},  t\in\set{\rm key,opp}),\\
    & \text{LCS}_{F_{t,t',a,a'}^{X,I}}^{\rm L}\left(Y_{a,X,I}^{b,\omega,1,t}\right) 
    \leq Y_{a',X,I}^{b,\omega,1,t'} \leq 
    \text{LCS}_{F_{t,t',a,a'}^{X,I}}^{\rm U}\left(Y_{a,X,I}^{b,\omega,1,t}\right),
    \quad (a\neq a'\in\set{0,1}, I\in\bar{I},  t\neq t'\in\set{\rm key,opp}),\\
    & 0\leq Y_{a,X,I}^{b,\omega,n} \leq 1
    \qquad (n\leq n_{\rm cut}, I\in\bar{I}),\\
    & 0\leq Y_{a,X,I}^{b,\omega,1,t} \leq 1
    \qquad (I\in\bar{I},t\in\set{\rm{key,opp}}),
\end{aligned}
\end{gather}
Here, $Y_{a,X,I}^{b,\omega,n}$ ($Y_{a,X,I}^{b,\omega,1,t}$), with $n=0,\dots,n_{\rm cut}$ ($t\in\set{\rm{key},\rm{opp}}$), denotes the probability that Bob observes the measurement outcome $b$ when he selects the $X$ basis and Alice emits the state $\rho_{a,\abasis,I_0}^{\omega,n}$ ($\ket*{\varphi_{a,\beta,I}^{\omega,t}}$); the quantity $Q_{a,X,I}^{b}$ denotes the probability that Bob observes the measurement outcome $b$ given that he measures the input state in the $X$ basis and Alice emits $\rho_{a,\abasis,I_0}^{\omega}$; and the new fidelities introduced in some constraints are defined as $F_{I,J}^{a,X,t}:=\abs*{\braket*{\varphi_{a,X,I}^{\omega,t}}{\varphi_{a,X,J}^{\omega,t}}}^2$ and $F_{t,t',a,a'}^{X,I}:=\abs*{\braket*{\varphi_{a,X,I}^{\omega,t}}{\varphi_{a',X,I}^{\omega,t'}}}^2$. Note that in the LP given in~\cref{eq:LP2appendix} we impose similarity constraints between the "key" eigenstate corresponding to a specific bit value and the "opp" eigenstate associated with the opposite bit value, as $\ket*{\varphi_{0,\beta,I}^{\omega,\rm{key}}}\approx\ket*{\varphi_{1,\beta,I}^{\omega,\rm{opp}}}$ and $\ket*{\varphi_{1,\beta,I}^{\omega,\rm{key}}}\approx\ket*{\varphi_{0,\beta,I}^{\omega,\rm{opp}}}$ for small $\omega$. 

Finally, an upper bound on the $X$-basis single-photon bit-error rate can be directly obtained as
\begin{equation}
    e_{X,I_0}^{\omega,1,\rm{key},{\rm U}}:=\frac{\Gamma_{X,I_0}^{\omega,1,\rm{key},{\rm U}}}{Y_{X,I_0}^{\omega,1,\rm{key},{\rm L}}},
\end{equation}
and, similarly to the main text, we compute the phase-error rate from the $X$-basis single-photon bit-error rate as
\begin{equation}
    e_{\rm ph}^{\omega,1,\rm{key},{\rm U}}:= G_{F_{ZX}'}^{\rm U}\left(e_{X,I_0}^{\omega,1,\rm{key},{\rm U}}\right),
\end{equation}
where
\begin{equation}\label{eq:fidelityZXappendix}
    F_{ZX}':=\left(1-\frac{1-\text{Re}\set{\braket*{\psi_{Z,I_0}^{\omega,\rm{key}}}{\psi_{X,I_0}^{\omega,\rm{key}}}}}{Y_{ZX\text{-coin}}^{\omega,1,\rm key,L}}\right)^2,
\end{equation}
$Y_{ZX\text{-coin}}^{\omega,1,\rm{key,L}}=(Y_{Z,I_0}^{\omega,1,\rm{key},{\rm L}}+Y_{X,I_0}^{\omega,1,\rm{key},{\rm L}})/2$, and $\ket*{\psi_{X,I_0}^{\omega,\rm{key}}}$ is defined in an analogous way to~\cref{eq:virtual_state_key} substituting the $Z$ basis by the $X$ basis. An important difference with the analysis introduced in~\cref{sec:Results} is that now the entangled states $\ket*{\psi_{\beta,I_0}^{\omega,\rm{key}}}$ do not depend on purifications of noisy mixed states, leading to an inner product in~\cref{eq:fidelityZXappendix} that is closer to one. 

\begin{figure}
    \centering
    \includegraphics[width=0.55\columnwidth]{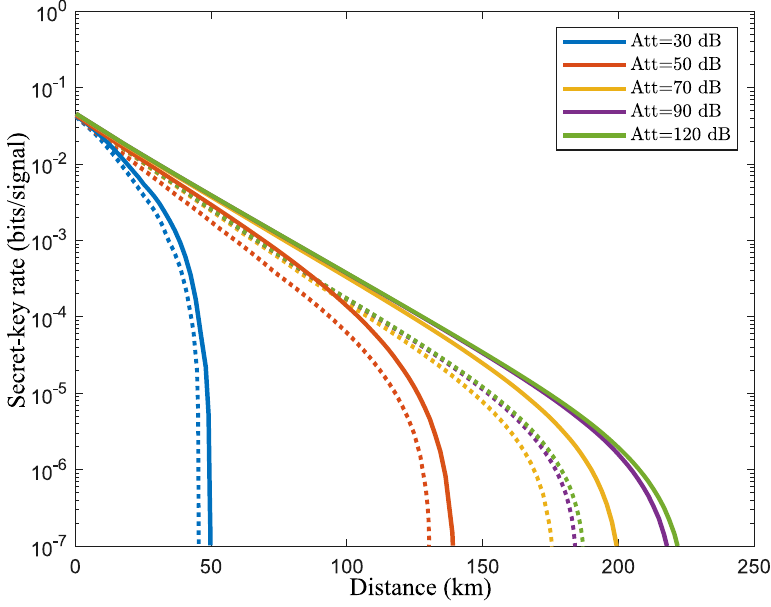}
    \caption{Asymptotic secret-key rate vs distance for the fully passive transmitter based on post-selection described in~\cref{sec:PassiveTransmitter}, for different values of the attenuation Att (in dB) introduced by Alice's IM to block the undesired pulses. Solid lines represent results obtained using the refined analysis presented in this appendix, while dotted lines correspond to the analysis introduced in~\cref{sec:security}. We consider a decoy-state BB84 protocol with three intensity settings. In the simulations, we numerically optimize over the parameters $\mu_{\rm max}$ and $\Delta\theta_Z$. For this comparison, we use the same experimental parameters as those in the main text.\label{fig:Passive_Tx_SKR_Comparison}}
\end{figure}

A comparison between the secret-key rate obtainable with the refined analysis introduced in this appendix and the one introduced in~\cref{sec:security} is shown in~\cref{fig:Passive_Tx_SKR_Comparison}. The refined analysis enhances the protocol's performance, particularly in scenarios where unwanted pulses are sufficiently attenuated. In fact, in the ideal setting with no information leakage, the refined analysis essentially matches ---precisely, marginally outperforms--- the one used in~\cite{lu2023experimental} in the limit $N\to\infty$.

\section{Channel model}\label{appendix:chanel_model}
For simplicity, in the simulations we assume that Bob uses an active BB84 receiver with two identical threshold detectors. That is, we assume that, when he selects the $Z$ basis, he performs measurements in the time bins $e$ and $l$, while when he selects the $X$ basis, he interferes $e$ and $l$ in a 50:50 BS and measures the output modes. Note that, since he does not measure in the odd time slots, the observed detection statistics do not depend on the information leakage.

\subsection{Observable quantities}
We start from the transmitted state given in~\cref{eq:state1rephrased}. After traveling through a channel with overall transmittance $\eta$ (which includes as well the efficiency of Bob's detectors), Bob receives
\begin{equation}
    \ket{\sqrt{\eta\mu}\cos(\theta/2)e^{i\phi_e}}_e \otimes \ket*{\sqrt{\eta\mu}\sin(\theta/2)e^{i(\phi+\phi_e)}}_l
\end{equation}
Clearly, the gain $Q_{\abit,\abasis,I}$ ---\textit{i.e.}, the probability that Bob observes a detection given that Alice post-selected the settings $\abit$, $\abasis$ and $I$--- is given by
\begin{equation}
    Q_{\abit,\abasis,I}=1-(1-p_d)^2 \frac{\langle e^{-\eta\mu}\rangle_{\Omega^{I}_{a,\beta}}}{\langle 1\rangle_{\Omega^{I}_{a,\beta}}},
\end{equation}
where $p_d$ is the dark-count rate of Bob's detectors.

For the bit-error probability we consider that double-click events are randomly assigned to a bit outcome~\cite{lutkenhaus1999estimates,lutkenhaus2000security}. 
This means that $ E_{0,Z,I}Q_{0,Z,I}=p_l^{0,Z,I}(1-p_e^{0,Z,I})+\frac{1}{2}p_l^{0,Z,I}p_e^{0,Z,I}$, where $p_m^{\abit,\abasis,I}$ is the probability that the detector is triggered in the time slot associated with the optical mode $m$ given that Alice selected the settings $\abit$, $\abasis$, and $I$. This leads to
\begin{equation}
    E_{0,Z,I}Q_{0,Z,I}=
    \frac{1}{2}Q_{0,Z,I}-\frac{1}{2}(1-p_d)^2\frac{\langle e^{-\eta\mu\sin^2(\theta/2)}-e^{-\eta\mu\cos^2(\theta/2)}\rangle_{\Omega^{I}_{0,Z}}}{\langle 1\rangle_{\Omega^{I}_{0,Z}}}.
\end{equation}
Moreover, we assume a symmetric channel that satisfies $E_{1,Z,I}=E_{0,Z,I}$. Regarding the $X$ basis, we have that
\begin{equation}
    E_{0,X,I}Q_{0,X,I}=\frac{1}{2}Q_{0,X,I}-\frac{1}{2}(1-p_d)^2\frac{\langle e^{\frac{\eta\mu}{2}[1-\sin(\theta)\cos(\phi)]}-e^{\frac{\eta\mu}{2}[1+\sin(\theta)\cos(\phi)]}\rangle_{\Omega^{I}_{0,X}}}{\langle 1\rangle_{\Omega^{I}_{0,X}}}.
\end{equation}

\subsection{Reference yields and bit-error probabilities}
To run the linear programs presented in the main text, we need some reference yields $\tilde{Y}_{\abasis,I,n}$ and $X$-basis bit-error probabilities $\tilde{\Gamma}_{X,I,n}$. In principle, all of these quantities could be numerically optimized to maximize the expected secret-key rate. However, as this is a computationally demanding task, we shall simply set them to their expected values given the channel model described above.

Specifically, for the reference yields, we simply take the values of the expected yields in an ideal BB84 scheme. That is, $\tilde{Y}_{\abasis,I}^{\omega,n}=1-(1-p_d)^2(1-\eta)^n$. 
As for the $n$-photon $X$-basis bit-error probabilities, if we define $p_{D_0,D_1}=\Pr[\text{click }D_0,\text{click }D_1]$, we have that
\begin{equation}
    \tilde{\Gamma}_{0,X,I}^{\omega,n}=p_{\bar{D_0},D_1}+\frac{1}{2}p_{D_0,D_1}=\text{Tr}[\rho_{X,I}^{\omega,n} \Pi_{\rm err}^{(n)}],
\end{equation}
where
\begin{equation}
\begin{split}
    \Pi_{\rm err}^{(n)}&
    =\Pi_{\bar{e}}^{(n)}-\Pi_{\bar{e},\bar{l}}^{(n)}+\frac{1}{2}\Pi_{e,l}^{(n)}
    \\
    \Pi_{\bar{x}}^{(n)}&=(1-p_d)\sum_{m=0}^{n}(1-\eta)^m\dyad{m}_x,
    \\
    \Pi_{\bar{e},\bar{l}}^{(n)}&=(1-p_d)^2\sum_{m=0}^{n}(1-\eta)^m\sum_{k=0}^n\dyad{k,m-k}_{e,l},
    \\
    \Pi_{e,l}^{(n)}&=\identity-\Pi_{\bar{e}}^{(n)}-\Pi_{\bar{l}}^{(n)}+\Pi_{\bar{e},\bar{l}}^{(n)}.
\end{split}
\end{equation}
Note that $\tilde{\Gamma}_{1,X,I}^{\omega,n}=\tilde{\Gamma}_{0,X,I}^{\omega,n}$, as we assume a symmetric channel.

%\newpage
%\input{Appendices/dev_Derivations}

%%%%%%%%%%%%%%%%%%%%%%%%%%%%%%%%%%%%%%%
\section*{References}
\bibliography{refs}
\end{document}